\documentclass{elsart}

\usepackage{graphicx}
\usepackage{amsmath}
\usepackage{amssymb}
\usepackage{amsbsy}
\usepackage{bm}

\numberwithin{equation}{section}
\journal{Nuclear Physics B}
\date{12 June 2006}

\newcommand{\intx}{\ensuremath{\int d^4 x}}

\newcommand{\sgn}{\ensuremath{\mathrm{sgn}}}
\newcommand{\intq}{\ensuremath{\int\frac{d^4 q}{(2\pi)^4}}}

\newcommand{\ppara}{\ensuremath{p_\parallel}}
\newcommand{\pperp}{\ensuremath{p_\perp}}
\newcommand{\qpara}{\ensuremath{q_\parallel}}
\newcommand{\qperp}{\ensuremath{q_\perp}}
\newcommand{\gammaperp}{\ensuremath{\gamma_\perp}}
\newcommand{\gammapara}{\ensuremath{\gamma_\parallel}}
\newcommand{\sumint}{\ensuremath{\sum\!\!\!\!\!\!\!\int}}
\newcommand{\rsumint}{\ensuremath{\sideset{}{'}\sum\!\!\!\!\!\!\!\!\!\int}}
\newcommand{\nn}{\nonumber}

\begin{document}
\begin{frontmatter}

\title{Gauge independent approach to chiral symmetry breaking in a
strong magnetic field}
\author[ud]{C. N. Leung} and \ead{leung@physics.udel.edu}
\author[ud,tku]{S.-Y. Wang\corauthref{cor}}\ead{sywang@mail.phys.tku.edu.tw}
\corauth[cor]{Corresponding author.}
\address[ud]{Department of Physics and Astronomy, University of Delaware,
Newark, Delaware 19716, USA}
\address[tku]{Department of Physics, Tamkang University, Tamsui,
Taipei 25137, Taiwan}

\begin{abstract}
The gauge independence of the dynamical fermion mass generated
through chiral symmetry breaking in QED in a strong, constant
external magnetic field is critically examined. We show that the
bare vertex approximation, in which the vertex corrections are
ignored, is a consistent truncation of the Schwinger-Dyson equations
in the lowest Landau level approximation. The dynamical fermion
mass, obtained as the solution of the truncated Schwinger-Dyson
equations evaluated on the fermion mass shell, is shown to be
manifestly gauge independent. By establishing a direct
correspondence between the truncated Schwinger-Dyson equations and
the 2PI (two-particle-irreducible) effective action truncated at the
lowest nontrivial order in the loop expansion as well as in the
$1/N_f$ expansion ($N_f$ is the number of fermion flavors), we argue
that in a strong magnetic field the dynamical fermion mass can be
reliably calculated in the bare vertex approximation.
\end{abstract}

\end{frontmatter}

\section{Introduction}

The study of chiral symmetry breaking has a long history, beginning
with the seminal work of Nambu and
Jona-Lasinio~\cite{Nambu,NJL1,NJL2}. Because of its simpler
structure, quantum electrodynamics (QED) has widely been used to
study chiral symmetry breaking in gauge
theories~\cite{JBW1,JBW2,MN1,MN2,FK,Miransky1,BLL1,BLL2,BLL3}. Being
inherently a nonperturbative phenomenon, the dynamical generation of
a fermion mass in gauge theories is usually studied with the help of
the Schwinger-Dyson equations truncated in certain schemes. An often
used truncation scheme is the rainbow (or quenched ladder)
approximation (for a review see Ref.~\cite{Miransky2}). It is
generally difficult to find a consistent truncation scheme such that
the resulting dynamical fermion mass is gauge (fixing) independent.
A common practice has been to forgo gauge independence in favor of
obtaining a nonperturbative solution. Nevertheless, the
demonstration of the gauge independence of physical quantities is of
fundamental importance in gauge
theories~\cite{Nielsen:1975fs,Kobes:1990dc}. In perturbation theory
the usual connection between the order of the loop expansion and
powers of the gauge coupling guarantees that physical quantities
calculated to a given order in the loop expansion are gauge
independent. In other words, as is well known, loop expansion is a
consistent expansion scheme in perturbation theory. Such a
connection however is lost in nonperturbative studies in which
contributions of a given order in the gauge coupling arise from
every order in the loop expansion. As a result, establishing the
gauge independence of physical quantities obtained in a
nonperturbative calculation (e.g., the Schwinger-Dyson equations) is
a highly nontrivial problem.  This is because an infinite subset of
diagrams arising from every order in the loop expansion has to be
resummed consistently, thus leading to potential gauge dependence of
physical quantities whenever not all relevant diagrams are accounted
for. Therefore, we emphasize that in gauge theories no truncation
schemes of the Schwinger-Dyson equations should be considered
consistent unless the gauge independence of physical quantities
calculated therein is unequivocally demonstrated.

The problem of chiral symmetry breaking in QED in a strong, constant
external magnetic field has received a lot of attention over the
past
decade~\cite{Gusynin:1994re,Gusynin:1994xp,Gusynin:1995gt,Gusynin:1995nb,Leung:1996qy,Hong:1996pv,Lee:1997zj,Ferrer:1998vw,Gusynin:1998zq,Gusynin:1999pq,Alexandre:2000nz,Alexandre:2001vu,Gusynin:2003dz,Kuznetsov:2002zq}.
It was originally motivated by a
proposal~\cite{prop1,prop2,prop3,prop4} to explain the correlated
$e^+e^-$ peaks observed in heavy ion collision
experiments~\cite{GSI1,GSI2}. The proposal posits that the $e^+e^-$
peaks are produced from the decay of a bound $e^+e^-$ system formed
in a new metastable phase of QED with broken chiral symmetry, which
is induced by the strong electromagnetic field present in the
neighborhood of the colliding heavy ions. Even though the $e^+e^-$
events were not observed in a later experiment~\cite{Argonne}, it is
still an interesting question to ask whether background fields can
induce chiral symmetry breaking. The result of the investigation
suggests that it may be relevant for studying the electroweak phase
transition in the early universe when a strong magnetic field was
present~\cite{Gusynin:1995gt,Gusynin:1995nb}. The methodology
developed also found interesting applications in the study of color
superconductivity in high density QCD~\cite{Ferrer:2005vd},
superconductivity and magnetic field induced phase transitions in
condensed matter
systems~\cite{Farakos:1998pn,Semenoff:1998bk,Liu:1998mg,Khveshchenko,Ferrer:2001fz},
as well as the properties of strongly magnetized astrophysical
media~\cite{Hong:1998ka,Elizalde:2004mw}.

In a recent article~\cite{Leung:2005xz}, we have found a consistent
truncation scheme for the case of chiral symmetry breaking in a
strong, constant external magnetic field.  We critically examined
the consistency and gauge independence of the bare vertex
approximation that has been extensively used in truncating the
Schwinger-Dyson equations to calculate the dynamical fermion mass
generated through chiral symmetry breaking in weakly coupled QED in
a strong magnetic field.  In particular, we showed that the bare vertex
approximation, in which the vertex corrections are ignored, is a
consistent truncation of the Schwinger-Dyson equations in the lowest
Landau level approximation.
The dynamical fermion mass, obtained as the solution of the
truncated Schwinger-Dyson equations evaluated \emph{on the fermion
mass shell}, is manifestly gauge independent.  This novel gauge
independent approach poses a serious question on the validity of the
results and conclusions obtained in earlier
studies~\cite{Gusynin:1995gt,Gusynin:1995nb,Leung:1996qy,Lee:1997zj,Gusynin:1998zq,Gusynin:1999pq,Alexandre:2000nz,Alexandre:2001vu,Gusynin:2003dz,Kuznetsov:2002zq}.
Most importantly, it leads to a consistent, nonperturbative
calculation of the dynamical fermion mass to leading order in the
gauge coupling, and allows one to identify the infinite subset of
diagrams that contribute to chiral symmetry breaking in a strong
magnetic field.

The purpose of this article is twofold. On the one hand, we present
the technical details of the results outlined in
Ref.~\cite{Leung:2005xz}. To begin with, we critically review the
properties of the so-called $E_p$ functions, first introduced by
Ritus~\cite{Ritus1,Ritus2} in the studies of QED in constant
external electromagnetic fields. In doing so, we correct a few
mistakes found in the
literature~\cite{Leung:1996qy,Lee:1997zj,Ferrer:1998vw,Ritus1,Ritus2,Elizalde:2000vz}.
We also derive the Ward-Takahashi identity in the bare vertex
approximation both within and beyond the lowest Landau level
approximation. This allows us to (i) correct a few oversights found
in Ref.~\cite{Ferrer:1998vw}, (ii) verify the conclusion obtained
therein on the Ward-Takahashi identity in the bare vertex
approximation and within the lowest Landau level approximation, and
(iii) show that the Ward-Takahashi identity in the bare vertex
approximation can be satisfied only within the lowest Landau level
approximation. By utilizing the $E_p$ function formalism and the
Ward-Takahashi identity in the bare vertex approximation, we prove
that the bare vertex approximation is a consistent truncation of the
Schwinger-Dyson equations in the lowest Landau level approximation.
In particular, we verify that (i) the truncated vacuum polarization
is transverse, (ii) the truncated fermion self-energy evaluated on
the fermion mass shell is manifestly gauge independent.

On the other hand, we provide a detailed comparison between our
results and those obtained in the
literature~\cite{Gusynin:1995gt,Gusynin:1995nb,Leung:1996qy,Lee:1997zj,Gusynin:1998zq,Gusynin:1999pq,Alexandre:2000nz,Alexandre:2001vu,Gusynin:2003dz,Kuznetsov:2002zq}.
Based on the gauge independent analysis presented in this article,
we argue that to a large extent the results and conclusions obtained
in those earlier studies can be attributed to inconsistent
truncation schemes, gauge dependent artifacts, or both. Along the
way this detailed comparison also establishes several important
aspects of the bare vertex approximation: (i) The assumption of a
momentum independent fermion self-energy, a consequence of the
Ward-Takahashi identity in the bare vertex approximation, is
reliable in the momentum region relevant to chiral symmetry breaking
in a strong magnetic field. (ii) While the extension to a momentum
dependent fermion self-energy violates the Ward-Takahashi identity
in the bare vertex approximation, the resultant gauge dependence of
the dynamical fermion mass is of higher order. (iii) In a strong
magnetic field the dynamical fermion mass can be reliably calculated
in the bare vertex approximation and is gauge independent up to
corrections of higher order. (iv) There exists a direct
correspondence between the Schwinger-Dyson equations truncated in
the bare vertex approximation and the nonperturbative approach based
on the 2PI (two-particle-irreducible) effective action truncated at
the lowest nontrivial order both in the loop expansion as well as in
the $1/N_f$ expansion ($N_f$ is the number of fermion flavors).

The rest of this article is organized as follows. In
Sec.~\ref{Sec:Ep} we critically review the properties of the Ritus
$E_p$ functions. In Sec.~\ref{Sec:SD} the Schwinger-Dyson equations
truncated in the bare vertex approximation are discussed, and the
Ward-Takahashi identity in the bare vertex approximation is derived
both within and beyond the lowest Landau level approximation. In
Sec.~\ref{Sec:Proof} we present the proof that the bare vertex
approximation is a consistent truncation of the Schwinger-Dyson
equations in the lowest Landau level approximation. The truncated
on-shell Schwinger-Dyson equations are numerically solved for
various numbers of fermion flavors, and an analytic fit for the
dynamical fermion mass is obtained. In Sec.~\ref{Sec:Comparison} we
compare our results to those found in the literature. We also
establish a direct contact with the 2PI effective action truncated
at the lowest nontrivial order in the loop expansion as well as in
the $1/N_f$ expansion. Finally, we present our conclusions in
Sec.~\ref{Sec:Conclusions}. In Appendix~\ref{Appendix:VC} we show
that the vacuum current vanishes identically in a constant external
magnetic field. Appendix~\ref{Appendix:2PI} provides a summary of
some exact relations derived from the 2PI effective action that are
useful to the discussion in the main text.

\section{Formalism: Ritus $\boldsymbol{E_p}$ functions}
\label{Sec:Ep}

In this section we critically review the properties of the $E_p$
functions, first introduced by Ritus in his seminal
work~\cite{Ritus1,Ritus2}. They form a complete set of Dirac
matrix-valued orthonormal functions and hence provide a convenient
formalism in the studies of QED in the presence of a constant
external magnetic
field~\cite{Leung:1996qy,Lee:1997zj,Ferrer:1998vw,Elizalde:2000vz}.
In doing so, a few mistakes and confusions found in the
literature~\cite{Leung:1996qy,Lee:1997zj,Ferrer:1998vw,Ritus1,Ritus2,Elizalde:2000vz}
are corrected and clarified.

We begin with the free field Dirac equation for a massless fermion
in the presence of a constant external magnetic field\footnote{In
this article we use the convention in which the metric has the
signature $g_{\mu\nu}=\mathrm{diag}(-1,1,1,1)$, the Dirac matrices
obey $\{\gamma^\mu,\gamma^\nu\}=-2g^{\mu\nu}$ and
$\gamma^5=i\gamma^0\gamma^1\gamma^2\gamma^3$.}
\begin{equation}
(\gamma\cdot\pi)\,\psi(x)=0,\label{Dirac}
\end{equation}
where $\pi_\mu=-i\partial_\mu-eA_\mu$ with $A_\mu$ being the
\emph{external} gauge field and $\psi$ is a Dirac spinor. Instead of
solving Eq.~\eqref{Dirac} directly, we show that there exists a
complete set of orthonormal functions, referred to in the literature
as the Ritus $E_p$ functions, such that in the basis spanned by the
$E_p$ functions Eq.~\eqref{Dirac} is rendered formally identical to
the Dirac equation for a massless fermion in the \emph{absence} of
external fields.

We will take the constant external magnetic field of strength $H$
in the $x_3$-direction with the corresponding vector potential
given by
\begin{equation}
A_\mu=(0,0,Hx_1,0),\label{Amu}
\end{equation}
where $\mu=0,1,2,3$. It is straightforward to verify that the
operators $(\gamma\cdot\pi)^2$, $\Sigma^3\equiv i\gamma^1\gamma^2$
and $\gamma^5$ constitute a maximal set of mutually commuting
operators.\footnote{Strictly speaking, this maximal set of mutually
commuting operators also contains the obvious operators
$-i\partial_0$, $-i\partial_2$ and $-i\partial_3$, which have been
omitted here for the sake of presentational simplicity.} The $E_p$
functions are constructed in terms of the simultaneous
eigenfunctions (eigenvectors) of these mutually commuting operators.

The eigenfunction equation for $(\gamma\cdot\pi)^2$ is given by
\begin{equation}
-(\gamma\cdot\pi)^2\,\phi_p(x)=p^2\,\phi_p(x),\label{eigeneq}
\end{equation}
where $\phi_p(x)$ is the eigenfunction corresponding to the
eigenvalue $p^2$. Note that the subscript $p$ in $\phi_p(x)$
denotes symbolically a set of quantum numbers yet to be
determined. In the chiral representation in which $\Sigma^3$ and
$\gamma^5$ are both diagonal with eigenvalues $\sigma=\pm 1$ and
$\chi=\pm 1$, respectively, the eigenfunctions $\phi_p(x)$ have the
general form
\begin{equation}
\phi_p(x)=E_{p\sigma}(x)\;\omega_{\sigma\chi},\label{eigenfunc1}
\end{equation}
where $\omega_{\sigma\chi}$ are bispinors which are the
eigenvectors of $\Sigma^3$ and $\gamma^5$. The scalar functions
$E_{p\sigma}(x)$ are found to be given by
\begin{equation}
E_{p\sigma}(x)=N(n)\,e^{i(p_0x^0+p_2x^2+p_3x^3)}
D_n(\rho),\label{eigenfunc2}
\end{equation}
where $N(n)=(4\pi|eH|)^{1/4}/\sqrt{n!}$ is a normalization factor,
and $D_n(\rho)$ denotes the parabolic cylinder
functions~\cite{Whittaker} with argument
$\rho=\sqrt{2|eH|}\,(x_1-p_2/eH)$ and nonnegative integer index
$n=0,1,2,\ldots$ given by
\begin{equation}
n=l+\frac{\sigma}{2}\,\sgn(eH)-\frac{1}{2}.\label{n}
\end{equation}
The nonnegative integer $l$ in Eq.~\eqref{n} labels the Landau
levels with $l=0$ being the lowest Landau level (LLL). Note that
in the absence of an external electric field, the scalar functions
$E_{p\sigma}(x)$ do not depend on the chirality $\chi$ (hence the
notation used).

The eigenvalue $p^2$ of $-(\gamma\cdot\pi)^2$ is identified as the
momentum squared of a massless fermion in an external magnetic
field. This can be understood by writing
$(\gamma\cdot\pi)^2=(\gammapara\cdot\pi_\parallel)^2
+(\gammaperp\cdot\pi_\perp)^2$ and noticing the following
properties:
\begin{align}
-(\gammapara \cdot \pi_\parallel)^2\,\phi_p(x)&\equiv
-(\gamma^0\pi_0+\gamma^3\pi_3)^2\,\phi_p(x)\nn\\
&=-(\pi_0^2-\pi_3^2)\,\phi_p(x)\nn\\
&=-(p_0^2-p_3^2)\,\phi_p(x)\nn\\
&\equiv\ppara^2\,\phi_p(x),\label{ppara}\\
-(\gammaperp \cdot \pi_\perp)^2\,\phi_p(x)&\equiv
-(\gamma^1\pi_1+\gamma^2\pi_2)^2\,\phi_p(x)\nn\\
&=(\pi_1^2+\pi_2^2-eH\Sigma^3)\,\phi_p(x)\nn\\
&=2|eH|l\,\phi_p(x)\nn\\
&\equiv p_\perp^2\,\phi_p(x),\label{pperp}
\end{align}
where, here and henceforth, the subscript $\parallel$ ($\perp$)
refers to the longitudinal: $\mu=0,3$ (transverse: $\mu=1,2$)
components. Hence, one finds
\begin{align}
p^2&=\ppara^2+\pperp^2\nn\\
&=-p_0^2+p_3^2+2|eH|l,
\end{align}
with $l$ being the quantum number of the quantized transverse
momentum squared and $\sqrt{2|eH|}$ the energy gap between adjacent
Landau levels (referred to henceforth as the Landau energy).

It is important to note that Eq.~\eqref{n}, together with the
nonnegativeness of $n$, imposes a constraint on the allowed value
of $\sigma$ when $l=0$, namely,
\begin{equation}
\sigma=\left\{
\begin{alignedat}{2}
&\sgn(eH)&\qquad&\text{for $l=0$,}\\
&\!\!\pm\sgn(eH)&&\text{for $l>0$.}
\end{alignedat}\right.\label{sigma}
\end{equation}
This is one of the subtle points that has been overlooked in the previous
literature~\cite{Leung:1996qy,Lee:1997zj,Ferrer:1998vw,Elizalde:2000vz}
that utilizes the Ritus $E_p$ functions. Physically, the constraint
on $\sigma$ when $l=0$ means that the spin of the LLL fermions is
always aligned parallel to the external magnetic field.

Following Ritus~\cite{Ritus1,Ritus2}, we construct the $E_p$
functions as the matrix of the simultaneous eigenfunctions
(eigenvectors) of the maximal set of mutually commuting operators
$\{(\gamma\cdot\pi)^2,\Sigma^3,\gamma^5\}$, viz,
\begin{equation}
E_p(x)=\sideset{}{'}\sum_{\sigma=\pm 1}
E_{p\sigma}(x)\,\Delta(\sigma),\label{Ep}
\end{equation}
where a prime on the summation symbol means that it is subject to
the constraint \eqref{sigma}, and
\begin{align}
\Delta(\sigma)&=\mathrm{diag}(\delta_{\sigma 1},
\delta_{\sigma -1},\delta_{\sigma 1},\delta_{\sigma -1})\nn\\
&=\frac{1}{2}(1+\sigma\,\Sigma^3).\label{Delta}
\end{align}
Note that the subscript $p$ in $E_p(x)$ denotes symbolically the
set of quantum numbers $\{p_0,p_2,p_3,l\}$.

It is straightforward to verify that (no summation over repeated
indices is implied)
\begin{equation}
\Delta(\sigma)+\Delta(-\sigma)=1,\quad
\Delta(\sigma)\,\Delta(\sigma')=\delta_{\sigma\sigma'}
\Delta(\sigma),\label{projection}
\end{equation}
hence the matrices $\Delta(\sigma)$ can be identified as the
projection operators on the fermion states with the spin parallel
($\sigma=1$) and antiparallel ($\sigma=-1$) to the external
magnetic field. Furthermore, $\Delta(\sigma)$ satisfy the
following important properties
\begin{gather}
[\Delta(\sigma),\gammapara^\mu]=0,\quad
\Delta(\sigma)\,\gammaperp^\mu=\gammaperp^\mu\,\Delta(-\sigma),
\label{DeltaProp}
\end{gather}
which are the key relations that will be frequently used in our
calculation.

Using the orthogonal property of the parabolic cylinder functions
\cite{Whittaker}
\begin{equation}
\int_{-\infty}^\infty d\rho\,D_{n'}(\rho)\,D_n(\rho)
=\sqrt{2\pi}\,n!\,\delta_{nn'},\label{Drho}
\end{equation}
we find that the $E_p$ functions are orthonormal and complete,
namely,
\begin{align}
\int d^4x\,\overline{E}_p(x)\,E_{p'}(x)
&=(2\pi)^4\,\widehat{\delta}^{(4)}(p-p')\,\Pi(l),\label{orthonormal}\\
\sumint\,d^4p\,E_p(x)\,\overline{E}_p(y)&=
(2\pi)^4\,\delta^{(4)}(x-y),\label{complete}
\end{align}
where $\overline{E}_p=\gamma^0E_p^\dagger\gamma^0$. Here we have
used the following shorthand notation
\begin{gather}
\Pi(l)=\left\{
\begin{alignedat}{2}
&\Delta[\sgn(eH)]&\qquad&\text{for $l=0$,}\\
&1&&\text{for $l>0$,}
\end{alignedat}\right.\\
\sumint\,d^4p=\sum^\infty_{l=0}
\int^\infty_{-\infty}dp_0\int^\infty_{-\infty}dp_2
\int^\infty_{-\infty}dp_3,\\
\widehat{\delta}^{(4)}(p-p')=\delta_{ll'}\,\delta(p_0-p'_0)\,
\delta(p_2-p'_2)\, \delta(p_3-p'_3).
\end{gather}
We emphasize that the orthonormal condition of the $E_p$ functions
for the LLL fermions (i.e., $l=0$) in Eq.~\eqref{orthonormal}
differs from that quoted in the
literature~\cite{Leung:1996qy,Lee:1997zj,Ferrer:1998vw,Elizalde:2000vz},
which is applicable \emph{only} for fermions occupying higher Landau
levels (i.e., $l>0$). The presence of the projection operator
$\Delta[\sgn(eH)]$ in the orthonormal condition of the $E_p$
functions for the LLL fermions has an important physical consequence
that is responsible for an effective dimensional reduction in the
dynamics of fermion pairing in a strong external magnetic
field~\cite{Gusynin:1994xp,Gusynin:1995nb}.

Since the $E_p$ functions form a complete set of orthonormal Dirac
matrix-valued functions, they can be used as a basis for the Hilbert
space of the (perturbative) asymptotic massless fermion states in
the presence of a constant external magnetic field. Throughout this
article we will refer to the space spanned by the $E_p$ functions
simply as the momentum space whenever no confusion may arise.

Explicit calculation shows that the $E_p$ functions satisfy an
important property
\begin{equation}
(\gamma\cdot\pi)\,E_p(x)=E_p(x)\,\gamma\cdot p,
\label{property1}
\end{equation}
where
\begin{equation}
p_\mu =(p_0,0,-\sgn(eH)\sqrt{2|eH|l},p_3).
\label{pmu}
\end{equation}
For notational simplicity, in Eq.~\eqref{property1} we have used the
\emph{same} notation $p$ to denote both symbolically the set of
quantum numbers $\{p_0,p_2,p_3,l\}$ on the left-hand side (LHS) in
$E_p(x)$, as well as literally the momentum $p_\mu$ on the
right-hand side (RHS) in $\gamma\cdot p$. However, we emphasize that
the quantum number $p_2$ should \emph{not} be confused with the
component of the momentum $p$ in the $x_2$-direction, which is
determined by the quantum number $l$ as can be seen from
Eq.~\eqref{pmu}.

We note that $\gamma\cdot\pi$ does not commute with $\Sigma^3$ and
$\gamma^5$, hence it is \emph{not} diagonal in the basis spanned by
the $E_p$ functions, as is evident from Eq.~\eqref{property1}.

Upon left multiplying Eq.~\eqref{property1} by
$\overline{E}_{p'}(x)$ and integrating over $x$, one obtains
\begin{align}
\int d^4x\,\overline{E}_{p'}(x)\,(\gamma\cdot\pi)\,E_p(x)
&=(2\pi)^4\,\widehat{\delta}^{(4)}(p-p')\,
\Pi(l)\,\gamma\cdot p.\label{Sinverse}
\end{align}
Eq.~\eqref{Sinverse} reveals clearly the advantage of the $E_p$
functions in the studies of QED in a constant external magnetic
field: When expressed in the momentum space spanned by the $E_p$
functions, the (bare) inverse propagator for fermions in a
constant external magnetic field is rendered \emph{formally}
identical to that in the absence of external fields (up to the
projection operator $\Delta[\sgn(eH)]$ that accounts for the spin
alignment of the LLL fermions).

We are now in a position to show that in the momentum space
spanned by the $E_p$ functions, Eq.~\eqref{Dirac} is rendered
formally identical to the Dirac equation for a massless fermion in
the absence of external fields. This is achieved by considering
the Dirac spinor in momentum space $\psi(p)$ defined by
\begin{equation}
\psi(x)=\sumint\,\frac{d^4p}{(2\pi)^4}\,E_{p}(x)\,\psi(p),\label{expansion}
\end{equation}
where the momentum argument $p$ in $\psi(p)$ is given by
Eq.~\eqref{pmu}. It follows from Eq.~\eqref{property1} that in
momentum space Eq.~\eqref{Dirac} reads
\begin{equation}
(\gamma\cdot p)\,\psi(p)=0,\label{Dirac1}
\end{equation}
which, as advertised in the beginning of this section, is formally
the Dirac equation for a massless fermion in the absence of external
fields, and whose solutions are well known. The condition that
Eq.~\eqref{Dirac1} has nontrivial solutions is given by $p^2=0$,
which is exactly the mass shell condition for a massless particle.
We emphasize that because the $E_p$ functions are not solutions of
the free field Dirac equation \eqref{Dirac}, they do \emph{not}
correspond to the (perturbative) asymptotic states of massless
fermions in the presence of a constant external magnetic field.
Nevertheless, as clearly displayed in Eqs.~\eqref{expansion} and
\eqref{Dirac1}, the latter can be expanded in the basis spanned by
the $E_p$ functions in terms of the usual Dirac (or Weyl) spinors
for massless fermions in the absence of external fields.

So far we have been considering only the noninteracting
theory for massless fermions in a constant external
magnetic field. Once the interaction with the (quantum) photon
field is taken into account, the fermion will receive radiative
corrections, which in turn give rise to the self-energy.
Consistent with Eq.~\eqref{property1}, the fermion self-energy in
coordinate space $\Sigma(x,x')$ has to satisfy a similar property,
\begin{equation}
\int d^4x'\,\Sigma(x,x')\,E_p(x')=E_p(x)\,\Sigma(p),
\label{property2}
\end{equation}
where $\Sigma(p)$ is a Dirac matrix-valued function (yet to be
determined by explicit calculation) with the momentum argument $p$
given by Eq.~\eqref{pmu}.

Following the same steps that lead to Eq.~\eqref{Sinverse}, one
finds the fermion self-energy in momentum space to be given by
\begin{align}
\Sigma(p,p')&=\int d^4x\,d^4x'\,
\overline{E}_p(x)\,\Sigma(x,x')\,E_{p'}(x')\nn\\
&=(2\pi)^4\,\widehat{\delta}^{(4)}(p-p')\,\Pi(l)\,\Sigma(p).
\label{SigmaP}
\end{align}
With Eqs.~\eqref{Sinverse} and \eqref{SigmaP}, one finds that the
full inverse fermion propagator in momentum space is given by [see
Eq.~\eqref{SDfermion}]
\begin{align}
G^{-1}(p,p')&=\int d^4x\,d^4x'\,
\overline{E}_p(x)\,G^{-1}(x,x')\,E_{p'}(x')\nn\\
&=(2\pi)^4\,\widehat{\delta}^{(4)}(p-p')\,\Pi(l)\,[\gamma\cdot
p+\Sigma(p)],\label{GinverseP}
\end{align}
which in turn implies that the full fermion propagator in momentum
space takes the form
\begin{align}
G(p,p')&=\int d^4x\,d^4x'\,
\overline{E}_p(x)\,G(x,x')\,E_{p'}(x')\nn\\
&=(2\pi)^4\,\widehat{\delta}^{(4)}(p-p')\,\Pi(l)\,
\frac{1}{\gamma\cdot p+\Sigma(p)}.\label{GP}
\end{align}
Eqs.~\eqref{SigmaP}-\eqref{GP}, together with the orthonormal
condition Eq.~\eqref{orthonormal} of the $E_p$ functions, imply that
in coordinate space one has
\begin{align}
\Sigma(x,x')&=\sumint\,\frac{d^4p}{(2\pi)^4}\,E_{p}(x)\,\Sigma(p)
\,\overline{E}_p(x'),\label{SigmaX}\\
G^{-1}(x,x')&=\sumint\,\frac{d^4p}{(2\pi)^4}\,E_{p}(x)\,[\gamma\cdot
p+\Sigma(p)]\,\overline{E}_p(x'),\label{GinverseX}\\
G(x,x')&=\sumint\,\frac{d^4p}{(2\pi)^4}\,E_{p}(x)
\frac{1}{\gamma\cdot p+\Sigma(p)}\,\overline{E}_p(x').\label{GX}
\end{align}
The set of equations \eqref{SigmaP}-\eqref{GX} constitutes the main
ingredients for the calculation presented in Secs.~\ref{Sec:SD} and
\ref{Sec:Proof}.

We emphasize that the fermion self-energy $\Sigma(p)$ in general is
\emph{not} a diagonal matrix, which is contrary to what has been
stated in the
literature~\cite{Leung:1996qy,Lee:1997zj,Ferrer:1998vw,Ritus1,Ritus2,Elizalde:2000vz}.
We shall now clarify this confusion. It was
argued~\cite{Leung:1996qy,Lee:1997zj,Ritus1,Ritus2,Elizalde:2000vz}
that in the presence of a constant magnetic field the fermion
self-energy operator $\widehat{\Sigma}$ is a Dirac matrix-valued
scalar function of the form
\begin{equation}
\widehat{\Sigma}=\widehat{\Sigma}
[\gamma\cdot\pi,(F_{\mu\nu}\pi^\nu)^2], \label{Sigmaop}
\end{equation}
where $F_{\mu \nu}=\partial_\mu A_\nu-\partial_\nu A_\mu$ is the
external field strength.  In order to highlight the fact that the
self-energy is in general an \emph{off-shell} quantity, we have
used the term self-energy operator instead of the usual term mass
operator, which can be misleading.

A comment here is in order. Based on symmetry
arguments~\cite{Ritus1,Ritus2}, for constant external fields the
self-energy operator $\widehat{\Sigma}$ in general depends on two
extra operators: $\gamma^5\,\widetilde{F}^{\mu\nu}F_{\mu\nu}$ and
$\sigma^{\mu\nu}F_{\mu\nu}$, where
$\widetilde{F}^{\mu\nu}=(1/2)\,\epsilon^{\mu\nu\sigma\rho}F_{\sigma\rho}$
and $\sigma^{\mu\nu}=(i/2)[\gamma^\mu,\gamma^\nu]$. The former
vanishes identically in the absence of an external electric field.
The latter renders either the fermion self-energy or the position of
the fermion pole (i.e., the fermion dispersion relation) explicitly
dependent on the sign of the external magnetic field strength, hence
is forbidden by a residual $Z_2$ symmetry of the rotational $O(3)$
symmetry that is broken in the presence of an external magnetic
field (see Sec.~\ref{Sec:SD} for a related discussion).

Since $(F_{\mu\nu}\pi^\nu)^2$ commutes with $(\gamma\cdot\pi)^2$,
the self-energy operator $\widehat{\Sigma}$ commutes with
$(\gamma\cdot\pi)^2$ and hence is
diagonal in the basis spanned by the eigenfunctions of
$(\gamma\cdot\pi)^2$. While this statement is correct in itself,
it does not imply that $\widehat{\Sigma}$ is diagonal in the basis
spanned by the $E_p$ functions. This can be understood as follows.
We note that even though $\widehat{\Sigma}$ commutes with
$(\gamma\cdot\pi)^2$, it does not commute with $\Sigma^3$ and
$\gamma^5$.  Therefore, $\widehat{\Sigma}$ is \emph{not} diagonal
in the basis spanned by the $E_p$ functions, which are constructed
in terms of the simultaneous eigenfunctions (eigenvectors) of the
maximal set of mutually commuting operators $\{(\gamma\cdot\pi)^2,
\Sigma^3,\gamma^5\}$.  This is precisely the same reason why the
operator $\gamma\cdot\pi$, in spite of commuting with
$(\gamma\cdot\pi)^2$, is not diagonal in the basis spanned by the
$E_p$ functions.

Last, but not least, we note that $\widehat{\Sigma}$ does \emph{not}
anticommute with $\gamma^5$. This in turn suggests, as one would
already have expected, dynamical chiral symmetry breaking in the
full theory of QED in the presence of a constant external magnetic
field. While can be confirmed only by explicit nonperturbative
calculation, the above symmetry arguments seem to indicate that
dynamical chiral symmetry breaking should be a generic phenomenon
independent of the number of fermion flavors. Indeed, it is one of
the goals of this article to verify in a gauge independent manner
that in a constant external magnetic field chiral symmetry is
dynamically broken regardless of the number of fermion flavors.

\section{Schwinger-Dyson equations and the bare vertex approximation}
\label{Sec:SD}

\begin{figure}[t]
\begin{center}
\includegraphics[width=3.5in,keepaspectratio=true,clip=true]{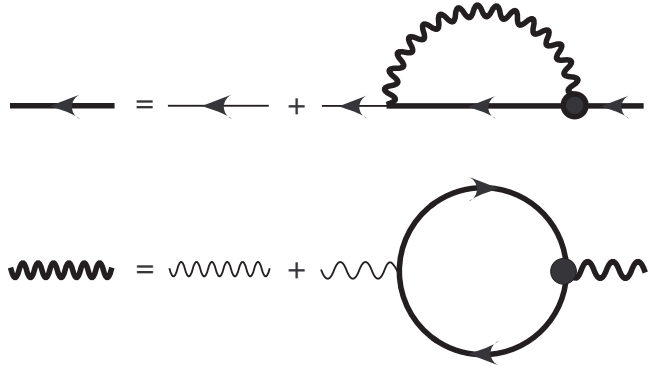}
\end{center}
\caption{The SD equations in QED in an external magnetic field. The
thin (heavy) line with an arrow denotes the bare (full) fermion
propagator in the presence of an external magnetic field. The thin
(heavy) wiggly line denotes the free (full) photon propagator. The
vertex with(out) a heavy dot denotes the full (bare) fermion-photon
interaction vertex.} \label{fig:SD}
\end{figure}

The Schwinger-Dyson (SD) equations in massless QED in an external
magnetic field are well-known in the literature (for reviews see
Refs.~\cite{Roberts:1994dr,Gitman}). The equations for the full
fermion propagator $G(x,y)$ are given by\footnote{We note that in
writing the fermion self-energy $\Sigma(x,y)$ as in
Eq.~\eqref{SDSigma}, it is implicitly assumed that the vacuum
current vanishes. In Appendix~\ref{Appendix:VC} we show that the
vacuum current does indeed vanish identically in a constant external
magnetic field.}
\begin{align}
G^{-1}(x,y)&=S^{-1}(x,y)+\Sigma(x,y),\label{SDfermion}\\
\Sigma(x,y)&=ie^2\intx' d^4y'
\,\gamma^\mu\,G(x,x')\Gamma^\nu(x',y,y')\,\mathcal{D}_{\mu\nu}(x,y'),
\label{SDSigma}
\end{align}
where $S(x,y)$ is the bare fermion propagator in the external gauge
field $A_\mu$ given by Eq.~\eqref{Amu}, $\Sigma(x,y)$ is the fermion
self-energy and $\Gamma^\nu(x,y,z)$ is the full vertex. The full
photon propagator $\mathcal{D}_{\mu\nu}(x,y)$ satisfies the
equations
\begin{align}
\mathcal{D}^{-1}_{\mu\nu}(x,y)&=D^{-1}_{\mu\nu}(x,y)+\Pi_{\mu\nu}(x,y),\\
\Pi_{\mu\nu}(x,y)&=-ie^2\,\mathrm{tr}\intx' d^4y'
\,\gamma_\mu\,G(x,x')\Gamma_\nu(x',y',y)\,G(y',x),\label{SDphoton}
\end{align}
where $D_{\mu\nu}(x,y)$ is the free photon propagator (defined in
covariant gauges) and $\Pi_{\mu\nu}(x,y)$ is the vacuum
polarization. The diagrammatical representation of the SD equations
\eqref{SDfermion}-\eqref{SDphoton} is depicted in Fig.~\ref{fig:SD}.

Since the dynamics of fermion pairing in a strong magnetic field is
dominated by the lowest Landau level
(LLL)~\cite{Gusynin:1995gt,Gusynin:1995nb,Leung:1996qy,Hong:1996pv,Lee:1997zj,Gusynin:1998zq,Gusynin:1999pq},
we will consider the propagation of, as well as radiative
corrections originating only from, fermions occupying the LLL.  This
is referred to as the lowest Landau level approximation (LLLA) in
the literature
\cite{Gusynin:1995gt,Gusynin:1995nb,Leung:1996qy,Hong:1996pv,Lee:1997zj,Ferrer:1998vw,Gusynin:1998zq,Gusynin:1999pq,Alexandre:2000nz,Alexandre:2001vu,Gusynin:2003dz,Kuznetsov:2002zq}.
Therefore, except where otherwise explicitly stated, the fermion
propagator and the fermion self-energy in the SD equations
\eqref{SDfermion}-\eqref{SDphoton} as well as in the rest of this
article will be taken to be those for the LLL fermions. For
notational simplicity, no separate notation will be introduced.

It is well known that the SD equations
\eqref{SDfermion}-\eqref{SDphoton} correspond to an infinite
hierarchy of integral equations. This is because the full vertex
(i.e., the three-point vertex function) $\Gamma^\mu$ depends on
four-point vertex functions which in turn depend on higher-point
vertex functions, and so on. In order to reduce the SD equations to
a closed system of integral equations that is tractable, a
truncation scheme of the SD equations needs to be employed by
truncating this infinite hierarchy at some point. Diagrammatically,
a truncation scheme of the SD equations corresponds to a resummation
of a selected infinite subset of diagrams arising from every order
in the loop expansion. The simplest truncation is done at the level
of the three-point vertex function by expressing the full vertex
$\Gamma^\mu$ in terms of the (bare or full) entities that already
appeared in the SD equations.

\begin{figure}[t]
\begin{center}
\includegraphics[width=3.5in,keepaspectratio=true,clip=true]{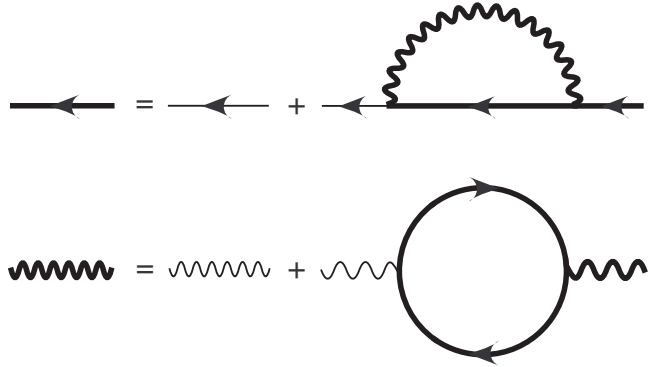}
\end{center}
\caption{The SD equations in the bare vertex approximation.}
\label{fig:SDBVA}
\end{figure}

To this end we will work in the bare vertex approximation (BVA),
in which the vertex corrections are completely ignored.
This is achieved by replacing the full vertex in the SD equations
\eqref{SDfermion}-\eqref{SDphoton} by the bare one, viz,
\begin{equation}
\Gamma^\mu(x,y,z)=\gamma^\mu\,\delta^{(4)}(x-z)\,\delta^{(4)}(y-z).
\label{BVA}
\end{equation}
The diagrammatical representation of the resultant SD equations
truncated in the BVA is depicted in Fig.~\ref{fig:SDBVA}. This
truncation is also known as the (improved) rainbow approximation and
has been employed extensively in the
literature~\cite{Gusynin:1995gt,Gusynin:1995nb,Leung:1996qy,Lee:1997zj,Gusynin:1998zq,Gusynin:1999pq,Alexandre:2000nz,Alexandre:2001vu,Gusynin:2003dz,Kuznetsov:2002zq}.
However, we emphasize that unlike what has usually been done in the
literature, here we will not confine ourselves to a particular gauge
(usually the Feynman
gauge)~\cite{Gusynin:1995gt,Gusynin:1995nb,Leung:1996qy,Lee:1997zj},
nor will we make the assumption that the BVA \eqref{BVA} is valid
only in a certain
gauge~\cite{Gusynin:1998zq,Gusynin:1999pq,Alexandre:2000nz,Alexandre:2001vu,Gusynin:2003dz,Kuznetsov:2002zq}.
Instead, it is our aim to prove that the BVA \eqref{BVA} is a
consistent truncation of the SD equations
\eqref{SDfermion}-\eqref{SDphoton} within the LLLA. The dynamical
fermion mass, obtained as the solution of the truncated SD equations
evaluated on the fermion mass shell, is manifestly gauge
independent. In the weak coupling regime that we consider, such a
gauge independent approach allows one to resum consistently an
infinite subset of diagrams that arise from every order in the loop
expansion and whose contributions are of leading order in the gauge
coupling, thus leading to a consistent and reliable calculation of
the dynamical fermion mass.

The main ingredient in the proof of the gauge independence of
physical quantities is the Ward-Takahashi (WT) identity satisfied by
the vertex and the inverse fermion propagator. It is
known~\cite{Miransky2,Atkinson:1979ut} that in ordinary QED the BVA
is not a consistent truncation of the SD equations because the
corresponding WT identity between the full fermion propagator and
the bare vertex in general cannot be fulfilled. The situation
however changes drastically in the presence of a strong, constant
external magnetic field as the motion of low-energy fermions is
restricted in directions perpendicular to the magnetic field,
leading to an effective dimensional reduction from $(3+1)$ to
$(1+1)$ in the dynamics of fermion pairing in a strong magnetic
field.

The WT identity in the BVA within the LLLA was first studied in
Ref.~\cite{Ferrer:1998vw}. It was shown that in order to satisfy the
WT identity in the BVA within the LLLA, the LLL fermion self-energy
in momentum space has to be a momentum independent constant.
However, due to a mistake in Ref.~\cite{Ferrer:1998vw} regarding the
Dirac matrix structure in the orthonormal condition of the Ritus
$E_p$ functions for the LLL fermions [see Eq.~\eqref{orthonormal}
and the discussion that follows], the calculation therein requires
further investigations. As we will see below, calculation performed
by using the correct orthonormal condition shows that the conclusion
obtained in Ref.~\cite{Ferrer:1998vw} on the WT identity in the BVA
remains valid. The reliability of such a momentum independent
approximation (in that the fermion self-energy has very weak
momentum dependence) and, consequently, of the WT identity in the
BVA has been verified within the LLLA in a certain gauge in the
momentum region relevant to the dynamics of fermion pairing in a
strong magnetic
field~\cite{Gusynin:1998zq,Gusynin:1999pq,Alexandre:2000nz,Alexandre:2001vu,Gusynin:2003dz}
(see further discussions in Sec.~\ref{Sec:Comparison}).

We now derive the WT identity in the BVA both \emph{within} and
\emph{beyond} the LLLA. This allows us to (i) correct a few
oversights found in Ref.~\cite{Ferrer:1998vw}, (ii) verify the
conclusion obtained therein on the WT identity in the BVA within
the LLLA, and (iii) show that the WT identity in the BVA can be
satisfied only within the LLLA. It is noted that in the rest of
this section the fermion propagator and the fermion self-energy
will in general \emph{not} be restricted to those for the LLL
fermions.

Following the standard procedure for deriving the WT in
QED~\cite{Itzykson,Brown,Ferrer:1998vw}, one
finds the WT identity in the BVA in the presence of an external
gauge field to be given by
\begin{equation}
-\partial_\mu^z[\delta^{(4)}(x-z)\,\delta^{(4)}(y-z)\,\gamma^\mu]
=i[\delta^{(4)}(y-z)-\delta^{(4)}(x-z)]\,G^{-1}(x,y),
\label{WT1}
\end{equation}
where it is noted that $G(x,y)$ is the full fermion propagator in
an external gauge field. Since the inverse fermion propagator in
the above equation does not depend on $z$, performing the Fourier
transform on the variable $z$ leads to
\begin{equation}
e^{-iq\cdot x}\,\delta^{(4)}(x-y)\,\gamma\cdot q=(e^{-iq\cdot x}
-e^{-iq\cdot y})\,G^{-1}(x,y),\label{WT2}
\end{equation}
where $q^\mu$ is the momentum carried by the photon. Upon left and
right multiplying Eq.~\eqref{WT2} by $\overline{E}_p(x)$ and
$E_{p'}(y)$, respectively, and integrating over $x$ and $y$, we
obtain the following identity
\begin{equation}
\int d^4x\,e^{-iq\cdot x}\,\overline{E}_p(x)\,\gamma\cdot q\,
E_{p'}(x)=\int d^4x\,d^4y\,(e^{-iq\cdot x}-e^{-iq\cdot y})\,
\overline{E}_p(x)\,G^{-1}(x,y)\,E_{p'}(y),\label{WT3}
\end{equation}
where the subscripts $p$ and $p'$ denote the sets
$\{p_0,p_2,p_3,l\}$ and $\{p'_0,p'_2,p'_3,l'\}$, respectively.

Let us first derive the WT identity in the BVA within the LLLA,
namely, $l=l'=0$ in Eq.~\eqref{WT3}. Using the properties of the
$E_p$ functions [see Eqs.~\eqref{Ep}, \eqref{Delta} and
\eqref{eigenfunc2}], one finds that the LHS of the WT identity
\eqref{WT3} is simplified to
\begin{equation}
\begin{aligned}
\int d^4x\,e^{-iq\cdot x}\,\overline{E}_p(x)\,\gamma\cdot
q\,E_{p'}(x)&=(2\pi)^4\,\delta^{(3)}(p+q-p')\,
e^{-q_\perp^2/4|eH|}\,e^{-iq_1(p_2+p'_2)/2eH}\\
&\quad\times\Delta[\sgn(eH)]\,\gammapara \cdot \qpara\,
\Delta[\sgn(eH)],
\end{aligned}
\label{LHS}
\end{equation}
where the transverse components $\gammaperp\cdot\qperp$ decouple due
to the second property of $\Delta(\sigma)$ in Eq.~\eqref{DeltaProp}.
Here $\qperp^2=q_1^2+q_2^2$ and
\begin{equation}
\delta^{(3)}(p+q-p')=\delta(p_0+q_0-p'_0)\,\delta(p_2+q_2-p'_2)
\,\delta(p_3+q_3-p'_3).
\end{equation}
Likewise, the RHS of the WT identity \eqref{WT3} can be simplified
to
\begin{align}
&\int d^4x\,d^4y \,(e^{-iq\cdot x}-e^{-iq\cdot y})\,
\overline{E}_p(x)\,G^{-1}(x,y)\,E_{p'}(y)\nn\\
&\quad\quad=(2\pi)^4\,\delta^{(3)}(p+q-p')\,e^{-q_\perp^2/4|eH|}
\,e^{-iq_1(p_2+p'_2)/2eH}\nn\\
&\quad\quad\quad\times\Delta[\sgn(eH)]\,[\gammapara \cdot
(p'-p)_\parallel + \Sigma(\ppara') - \Sigma(\ppara)] \,
\Delta[\sgn(eH)],\label{RHS}
\end{align}
where used has been made of Eq.~\eqref{GinverseX}, and the momenta
$p$ and $p'$ on the RHS in the square brackets are given by
$p_\mu=(p_0,0,0,p_3)$ and $p'_\mu=(p'_0,0,0,p'_3)$, respectively
[see the remarks after Eq.~\eqref{pmu}]. Upon collecting the above
results and using the delta functions to express $\ppara'$ in terms
of $(p+q)_\parallel$, we can rewrite the WT identity \eqref{WT3} in
the following compact form
\begin{equation}
\Delta[\sgn(eH)]\,\bigl\{\gammapara\cdot\qpara-[\gammapara\cdot
(p+q)_\parallel + \Sigma(\ppara+\qpara)]+[\gammapara\cdot\ppara +
\Sigma(\ppara)]\bigr\} \, \Delta[\sgn(eH)]=0.
\label{WT4}
\end{equation}
In the above expression, we have purposely
kept the tree-level terms, which cancel among themselves, to
highlight the most general form of the WT identity in the BVA
within the LLLA.

To proceed further, one needs to know the Dirac matrix structure of
the fermion self-energy $\Sigma(p)$. Based on symmetry arguments~\cite{Ferrer:1998vw},
for a constant external magnetic field $\Sigma(p)$ takes the
following form
\begin{equation}
\Sigma(p)=A_\parallel(p)\,\gammapara\cdot\ppara+
A_\perp(p)\,\gammaperp\cdot\pperp+B(p),\label{SigmaComp}
\end{equation}
where $A_{\parallel,\perp}(p)$ and $B(p)$ are functions of the
longitudinal and transverse momentum squared as well as the
magnitude of the external magnetic field, yet to be determined by
explicit calculation.

A comment here is in order. It was shown in
Ref.~\cite{Ferrer:1998vw} that for a constant external magnetic
field along the $x_3$-direction, the most general structure of
$\Sigma(p)$ contains two extra terms that are proportional to
$H\Sigma^3$, where $H$ is the strength of the constant external
magnetic field. Detailed analysis indicates that such terms, if
present, lead to either the fermion self-energy or the position of
the fermion pole being explicitly dependent on the sign of the
external magnetic field strength $H$, hence are forbidden by a
residual $Z_2$ symmetry (i.e., $H\to -H$) of the rotational $O(3)$
symmetry that is broken in the presence of an external magnetic
field (see Sec.~\ref{Sec:Ep} for a related discussion).

Upon substituting Eq.~\eqref{SigmaComp} into Eq.~\eqref{WT4}, one
finds that the transverse components proportional to
$\gammaperp\cdot\pperp$ in Eq.~\eqref{SigmaComp} decouple from the
WT identity. Furthermore, the requirement that the WT identity is
fulfilled for \emph{arbitrary} momenta $p$ and $q$ regardless of the
sign of $H$, leads to the conclusion that $\Sigma(\ppara)$ has to be
a \emph{momentum independent constant}.

Before ending this section, let us derive the WT identity in the BVA
but \emph{beyond} the LLLA, where one of the fermions occupies the
lowest Landau level while the other fermion occupies a higher Landau
level.  In other words, we consider Eq.~\eqref{WT3} with $l=0$ and
$l'>0$. We shall show that
the WT identity in the BVA can be satisfied only within the LLLA.
This is an important conclusion that allows for an unequivocal
refutation of the results and conclusions obtained in
Ref.~\cite{Kuznetsov:2002zq}, which can be attributed to an
inconsistent truncation of the SD equations beyond the LLLA (see
Sec.~\ref{Sec:Comparison} for details).

Following similar steps as above, one can rewrite the WT identity
\eqref{WT3} in the BVA but beyond the LLLA in the following compact
form ($l'>0$)
\begin{align}
\Delta[\sgn(eH)]\,&\bigl\{K_{l'}(\qperp^2,\varphi)\bigl[\gammapara\cdot\qpara
- \gamma \cdot p' +
\gammapara\cdot\ppara-\Sigma(p')+\Sigma(\ppara)\bigr]\nn\\
&+K_{l'-1}(\qperp^2,\varphi)\,\gammaperp\cdot\qperp\bigr\}=0,\label{WT5}
\end{align}
where use has been made of the properties of $\Delta(\sigma)$ in
Eq.~\eqref{DeltaProp} as well as the Dirac matrix structure of the
fermion self-energy in Eq.~\eqref{SigmaComp}, and the momenta $p$,
$p'$ and $q$ are respectively given by [see the remarks after
Eq.~\eqref{pmu}]
\begin{align}
p_\mu&=(p_0,0,0,p_3),\nn\\
p'_\mu&=(p'_0,0,-\sgn(eH)\sqrt{2|eH|l'},p'_3),\\
q_\mu&=(p'_0-p_0,q_1,p'_2-p_2,p'_3-p_3).\nn
\end{align}
In the above expression, we have use the shorthand notation
\begin{align}
K_{l'}(\qperp^2,\varphi)&=\frac{1}{\sqrt{l'!}}\;
\exp[-i\,\sgn(eH)\,l'\varphi]\Biggl[i\,\sgn(eH)\biggl(\frac{\qperp^2}{2|eH|}\biggr)^{1/2}
\Biggr]^{l'},\label{Kl}
\end{align}
where $\varphi=\arctan(q_2/q_1)$. It is an easy exercise to verify
that for either a momentum independent or a momentum dependent
fermion self-energy, Eq.~\eqref{WT5} cannot be satisfied for
arbitrary momenta $p$ and $p'$ regardless of the sign of $H$. This
result is not unexpected because, as is evident from
Eq.~\eqref{WT5}, there is no longer an effective dimensional
reduction in the dynamics of fermion pairing beyond the LLLA. The
situation is then similar to that in ordinary QED, in which it is
known that the BVA is not a consistent truncation of the SD
equations~\cite{Miransky2,Atkinson:1979ut}.  Consequently, we
conclude that (i) the WT identity in the BVA can be satisfied only
within the LLLA with a momentum independent fermion self-energy;
(ii) in order to go beyond the LLLA one has to use truncation
schemes of the SD equations that consistently account for vertex
corrections. To the best of our knowledge, such a consistent
truncation of the SD equations beyond the LLLA has not appeared in
the literature.

\section{On-shell gauge independence in the bare vertex approximation}
\label{Sec:Proof}

As per the WT identity in the BVA and within the LLLA \eqref{WT4},
we can write the self-energy for the LLL fermion as
\begin{equation}
\Sigma(\ppara)=m(\xi),
\end{equation}
where $m(\xi)$ is a momentum independent \emph{but} gauge
dependent constant, with $\xi$ being the gauge fixing parameter in
covariant gauges. It is important to note that $m(\xi)$ depends
implicitly on $\xi$ through the full photon propagator
$\mathcal{D}_{\mu\nu}$ in Eq.~\eqref{SDSigma}. We emphasize that
because of its implicit $\xi$-dependence, $m(\xi)$ should \emph{not} be
taken for granted to be the dynamical fermion mass, which is a
gauge independent physical quantity.

We now begin the proof that the BVA is a consistent truncation of
the SD equations \eqref{SDfermion}-\eqref{SDphoton}, in which
$m(\xi)$ is $\xi$-independent and hence can be identified
unambiguously as the dynamical fermion mass, \emph{if and only if}
the truncated SD equation for the fermion self-energy is evaluated
on the fermion mass shell.

We first recall that, as proved in Ref.~\cite{Kobes:1990dc}, in
gauge theories the singularity structures (i.e., the positions of
poles and branch singularities) of gauge boson and fermion
propagators are gauge independent when all contributions of a given
order of a systematic expansion scheme are accounted for.
Consequently, this means the dynamical fermion mass has to be
determined by the pole of the full fermion propagator obtained in a
consistent truncation scheme.

In momentum space the full propagator for the LLL fermion is given
by [see Eq.~\eqref{GP}]
\begin{equation}
G(\ppara)=\frac{1}{\gammapara\cdot\ppara+\Sigma(\ppara)}\,
\Delta[\sgn(eH)],\label{G}
\end{equation}
where, as noted in Sec.~\ref{Sec:Ep}, $\Delta[\sgn(eH)]$ is the
projection operator on the fermion states with the spin parallel to
the external magnetic field. Assume for the moment that the BVA is a
consistent truncation of the SD equations in the LLLA, such that the
position of the pole of the LLL fermion propagator $G(\ppara)$ is
gauge independent. It follows that in accordance with the WT
identity in the BVA \eqref{WT4} or, equivalently, the momentum
independence of the LLL fermion self-energy, we have
\begin{equation}
\Sigma(\ppara)=\Sigma(\ppara^2=-m^2)=m,\label{Sigma}
\end{equation}
where $m$ is the \emph{gauge independent, physical dynamical fermion
mass}, yet to be determined by solving the truncated SD equations
self-consistently. With the LLL fermion self-energy given by
Eq.~\eqref{Sigma}, the WT identity in the BVA \eqref{WT4} reduces to
a tree-like form
\begin{equation}
\gammapara\cdot\qpara=(\gammapara\cdot\ppara+m)-
[\gammapara\cdot(p-q)_\parallel+m].\label{WT}
\end{equation}

What remains to be verified in our proof that the BVA is a
consistent truncation of the SD equations in the LLLA is the
following statements: (i) the truncated vacuum polarization is
transverse, (ii) the truncated fermion self-energy when evaluated on
the fermion mass shell ($\ppara^2=-m^2$) is manifestly gauge
independent. We highlight that the fermion mass shell condition is
one of the most important points that has gone unnoticed in the
literature, where the truncated fermion self-energy used to be
evaluated off the fermion mass shell at, say,
$\ppara^2=0$~\cite{Gusynin:1995gt,Gusynin:1995nb,Leung:1996qy,Lee:1997zj,Gusynin:1998zq,Gusynin:1999pq,Alexandre:2000nz,Alexandre:2001vu,Gusynin:2003dz,Kuznetsov:2002zq}.

Let us first consider the vacuum polarization, which in the BVA
reads
\begin{equation}
\Pi^{\mu\nu}(x,y)=-ie^2\,\mathrm{tr}\,\gamma^\mu\,G(x,y)\,\gamma^\nu\,G(y,x),
\label{Pixy}
\end{equation}
where the LLL fermion propagator $G(x,y)$ is given by
Eq.~\eqref{GX} with the summation over $l$ restricted to the $l=0$
term and, as per the WT identity in the BVA \eqref{WT4}, a
momentum independent self-energy given by Eq.~\eqref{Sigma}. In
other words, one has
\begin{equation}
G(x,y)=\rsumint\;\frac{d^4p}{(2\pi)^4}\,E_{p}(x)
\frac{1}{\gammapara\cdot\ppara+m}\,\overline{E}_p(y),\label{GLLL}
\end{equation}
where a prime on the summation symbol means the summation over $l$
is restricted to the $l=0$ term. Taking the Fourier transform of
Eq.~\eqref{Pixy} and using the properties in
Eq.~\eqref{DeltaProp}, we find the vacuum polarization in momentum
space to be given by
\begin{align}
\Pi^{\mu\nu}(q)&=-\frac{ie^2}{2\pi}\,|eH|\,
\exp\biggl(-\frac{q_\perp^2}{2|eH|}\biggr)\,
\mathrm{tr}\int\frac{d^2p_\parallel}{(2\pi)^2}\nn\\
&\quad\times\gamma^\mu_\parallel\,\frac{1}{\gammapara\cdot\ppara+m}\,
\gamma^\nu_\parallel\,\frac{1}{\gammapara\cdot(p-q)_\parallel+m}
\Delta[\sgn(eH)].\label{Pi}
\end{align}
The presence of $\Delta[\sgn(eH)]$ in Eq.~\eqref{Pi} is a
consequence of the LLLA, which, as explicitly displayed in
Eq.~\eqref{Pi}, leads to an effective dimensional reduction from
$(3+1)$ to $(1+1)$ as the LLL fermions couple only to the
longitudinal components of the photon field.

The WT identity in the BVA \eqref{WT} guarantees that the vacuum
polarization $\Pi_{\mu\nu}(q)$ is transverse, viz,
\begin{equation}
q^\mu \Pi_{\mu\nu}(q)=0.
\end{equation}
Explicit calculation in dimensional regularization shows that the
$1/\epsilon$ pole corresponding to an ultraviolet logarithmic
divergence cancels, leading to
\begin{equation}
\Pi^{\mu\nu}(q)=\Pi(\qpara^2,\qperp^2)
\biggl(g_\parallel^{\mu\nu}-\frac{q^\mu_\parallel
q^\nu_\parallel}{\qpara^2}\biggr).
\end{equation}
This in turn implies that the full photon propagator takes the
following form in covariant gauges ($\xi=1$ is the Feynman gauge):
\begin{equation}
\mathcal{D}^{\mu\nu}(q)=\frac{1}{q^2+\Pi(\qpara^2,\qperp^2)}
\biggl(g_\parallel^{\mu\nu}-\frac{q^\mu_\parallel
q^\nu_\parallel}{\qpara^2}\biggr)+\frac{g_\perp^{\mu\nu}}{q^2}
+\frac{q^\mu_\parallel q^\nu_\parallel}{q^2
\qpara^2}+(\xi-1)\frac{1}{q^2}\frac{q^\mu q^\nu}{q^2}. \label{D}
\end{equation}
The polarization function $\Pi(\qpara^2,\qperp^2)$ is given by
\begin{equation}
\Pi(\qpara^2,\qperp^2)=\frac{2\alpha}{\pi}\,
|eH|\,\exp\biggl(-\frac{q_\perp^2}{2|eH|}\biggr)
F\biggl(\frac{\qpara^2}{4m^2}\biggr),\label{Piq}
\end{equation}
where $\alpha=e^2/4\pi$ is the fine-structure constant and
\begin{equation}
F(u)=1-\frac{1}{2u\sqrt{1+1/u}}\,\log\frac{\sqrt{1+1/u}+1}{\sqrt{1+1/u}-1}.
\end{equation}
Our result for $\Pi(\qpara^2,\qperp^2)$ agrees with those obtained
in the literature~\cite{Gusynin:1995nb,Hong:1996pv,Calucci:1993fi}.
It is worth noticing that the result in Ref.~\cite{Hong:1996pv} is
valid only in the limit $m \to 0$.

\begin{figure}[t]
\begin{center}
\includegraphics[width=4.0in,keepaspectratio=true,clip=true]{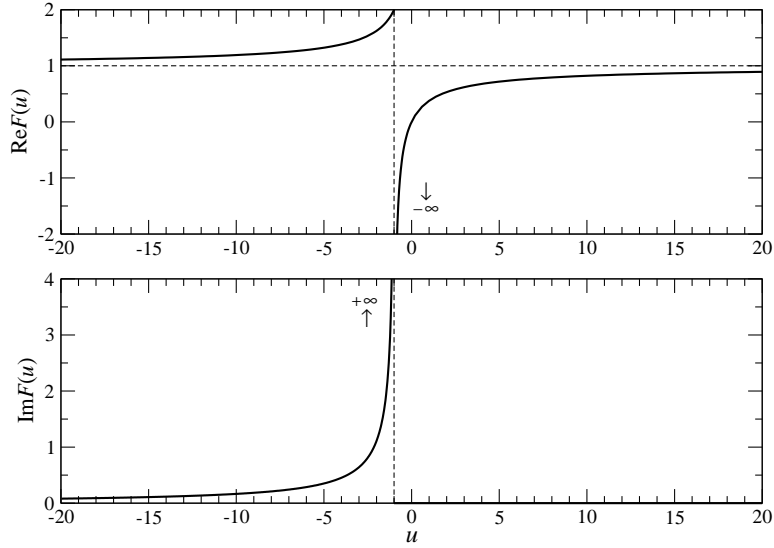}
\end{center}
\caption{Plot of the real and imaginary parts of $F(u)$ as a
function of $u$.} \label{fig:fofu}
\end{figure}

A detailed analysis of the analytic structure of the function
$F(u)$ shows that $F(u)$ has an imaginary part for $u<-1$, viz,
\begin{equation}
\mathrm{Im}F(u)=-\frac{\pi}{2u\sqrt{1+1/u}}\,\theta(-u-1),
\end{equation}
where $\theta(x)$ is the Heaviside step function. Hence the
polarization function $\Pi(\qpara^2,\qperp^2)$ is complex for
$\qpara^2<-4m^2$, with an imaginary part given by
\begin{equation}
\mathrm{Im}\Pi(\qpara^2,\qperp^2)=-\frac{2\alpha}{\pi}\,
|eH|\,\sgn(q_0)\,\exp\biggl(-\frac{q_\perp^2}{2|eH|}\biggr)\,
\mathrm{Im}F\biggl(\frac{\qpara^2}{4m^2}\biggr).\label{ImPi}
\end{equation}
We note that the extra factor of $-\sgn(q_0)$ is due to the analytic
continuation to the complex $q^0$-plane, namely, $q^0\to
q^0+i\epsilon$ with $\epsilon\to 0^+$. As will be seen momentarily,
the property that $\mathrm{Im}\Pi(\qpara^2,\qperp^2)$ is an odd
function of $q_0$ has an important consequence on the dynamical
fermion mass. The real and imaginary parts of $F(u)$ are depicted in
Fig.~\ref{fig:fofu}. The real part of $F(u)$ has the following
asymptotic behavior:
\begin{equation}
\mathrm{Re}F(u)=\left\{
\begin{alignedat}{2}
&\frac{2u}{3}+\mathcal{O}(u^2)&&\text{for $|u|\ll 1$,}\\
&1-\frac{1}{2u}\log(4|u|)+\mathcal{O}\biggl(\frac{1}{u^2}\biggr)&\quad&
\text{for $|u|\gg 1$.}
\end{alignedat}\right.
\end{equation}
The polarization effects modify drastically the propagation of
virtual photons in a constant external magnetic field. While photons
of momenta $|\qpara^2|\ll m^2$ remain unscreened, photons of momenta
$m^2\ll|\qpara^2|\ll|eH|$ and $\qperp^2\ll |eH|$ are screened with a
characteristic length $L=(2\alpha|eH|/\pi)^{-1/2}$. It is noted that
the upper limit $|eH|$ in the range of $|\qpara^2|$ is due to the
LLLA. As will be seen in Sec.~\ref{Sec:Comparison}, this screening
effect renders the rainbow
approximation~\cite{Gusynin:1995gt,Gusynin:1995nb,Leung:1996qy,Lee:1997zj},
in which the bare vertex as well as the free photon propagator are
used, completely unreliable in this problem.

We next consider the fermion self-energy, which in the BVA reads
\begin{equation}
\Sigma(x,y)=ie^2\,\gamma^\mu\,G(x,y)\,\gamma^\nu\,\mathcal{D}_{\mu\nu}(x,y),
\label{Sigmaxy}
\end{equation}
where $G(x,y)$ is the LLL fermion propagator given by
Eq.~\eqref{GLLL} and $\mathcal{D}_{\mu\nu}(x,y)$ is the full
photon propagator
\begin{equation}
\mathcal{D}_{\mu\nu}(x,y)=\int\frac{d^4q}{(2\pi)^4}\,e^{iq\cdot(x-y)}\,
\mathcal{D}_{\mu\nu}(q),\label{Dxy}
\end{equation}
with $\mathcal{D}_{\mu\nu}(q)$ given by Eq.~\eqref{D}.

Following similar steps in Sec.~\ref{Sec:SD} in the derivation of
the WT identity by left and right multiplying Eq.~\eqref{Sigmaxy},
respectively, by $\overline{E}_p(x)$ and $E_{p'}(y)$ corresponding
to the LLL fermion (i.e., $l=l'=0$), and integrating over $x$ and
$y$, we find that in momentum space the fermion self-energy
evaluated on the fermion mass shell is given by
\begin{eqnarray}
m\,\Delta[\sgn(eH)]&=&ie^2\intq\,\exp\biggl(-\frac{q_\perp^2}{2|eH|}\biggr)
\gammapara^\mu\,\frac{1}{\gammapara\cdot(p-q)_\parallel+m}
\,\gammapara^\nu\nn\\
&&\times\mathcal{D}_{\mu\nu}(q)\,\Delta[\sgn(eH)]\biggr\lvert_{\ppara^2=-m^2},
\label{m}
\end{eqnarray}
where use has been made of Eq.~\eqref{DeltaProp}. The presence of
$\Delta[\sgn(eH)]$ in \eqref{m} is again a consequence of the
LLLA.

We note that the RHS of Eq.~\eqref{m} appears to contain a
gauge dependent part that arises from the gauge
dependent term in $\mathcal{D}_{\mu\nu}(q)$ [see Eq.~\eqref{D}].
The WT identity in the BVA \eqref{WT} guarantees that this
\emph{would-be} gauge dependent contribution to the fermion self-energy
(denoted symbolically as $\Sigma_\xi$) is proportional to
$(\gammapara \cdot \ppara + m)$.
An explicit calculation shows that $\Sigma_\xi$ is
given by
\begin{align}
\Sigma_\xi&=\alpha\,(\xi-1)\,(\gammapara\cdot\ppara+m)\int_0^1
dx\,(1-x) \int\frac{d^2\qperp}{(2\pi)^2}\,
\exp\biggl(-\frac{q_\perp^2}{2|eH|}\biggr)\nn\\
&\quad\times\frac{(1-x)\qperp^2-xm(\gammapara\cdot\ppara-m)}{[(1-x)
\qperp^2+x(1-x)\ppara^2+xm^2]^2}\,\Delta[\sgn(eH)]. \label{mxi}
\end{align}
Since $\Sigma_\xi$ is proportional to $(\gammapara\cdot\ppara+m)$,
it vanishes identically on the fermion mass shell $\ppara^2=-m^2$
or, equivalently, $\gammapara\cdot\ppara+m=0$. This, together with
the transversality of the vacuum polarization, completes our proof
that the BVA is a consistent truncation of the SD equations.
Consequently, the dynamical fermion mass, obtained as the solution
of the truncated SD equations evaluated on the fermion mass shell,
is manifestly gauge independent.

Having proved the gauge independence of the dynamical fermion mass
in the BVA, we are now ready to find $m$ by solving Eq.~\eqref{m}
self-consistently. Note that the transverse components in
$\mathcal{D}^{\mu\nu}(q)$ decouple in the LLLA. Following the same
argument given above in the proof of the on-shell gauge
independence of the fermion self-energy, one can verify fairly easily
that contributions from
the longitudinal components in $\mathcal{D}^{\mu\nu}(q)$
proportional to $q^\mu_\parallel q^\nu_\parallel/\qpara^2$ are also
proportional to $(\gammapara \cdot \ppara + m)$, hence they vanish
identically on the fermion mass shell. Therefore, only the first
term in $\mathcal{D}^{\mu\nu}(q)$ proportional to
$g_\parallel^{\mu\nu}$ contributes to the on-shell SD equation
\eqref{m}. Consequently, the Dirac matrix structures on both sides
of Eq.~\eqref{m} are consistent. With these remarks we obtain
\begin{align}
m&=-ie^2\int\frac{d^4q}{(2\pi)^4}\frac{2m}{(p-q)_\parallel^2+m^2}\,
\exp\biggl(-\frac{q_\perp^2}{2|eH|}\biggr)\nn\\
&\quad\times \frac{q^2+\mathrm{Re}\Pi(\qpara^2,\qperp^2)-
i\mathrm{Im}\Pi(\qpara^2,\qperp^2)}{[q^2+\mathrm{Re}\Pi(\qpara^2,\qperp^2)]^2
+[\mathrm{Im}\Pi(\qpara^2,\qperp^2)]^2}\biggr\lvert_{\ppara^2=-m^2}.
\label{m1}
\end{align}
Since $\mathrm{Im}\Pi(\qpara^2,\qperp^2)$ is an odd function of
$q_0$ [see Eq.~\eqref{ImPi}], the imaginary part of the above
integral vanishes identically. As a consequence, the dynamical
fermion mass $m$ determined by the on-shell SD equation \eqref{m1}
is a real-valued quantity as it should be. Anticipating that the
dynamical fermion mass is much less than the Landau energy, i.e.,
$m^2\ll|eH|$, we find that the real part of the above integral is
dominated by the region of momentum $m^2\ll|\qpara^2|\ll|eH|$ and
$\qperp^2\ll|eH|$. Therefore, the imaginary part of the polarization
function $\mathrm{Im}\Pi(\qpara^2,\qperp^2)$ in the denominator in
Eq.~\eqref{m1} can be neglected. Using the mass shell condition
$\ppara^\mu=(m,0)$ that corresponds to a LLL fermion at rest and
performing a Wick rotation to Euclidean space, we find that
Eq.~\eqref{m1} in Euclidean space reads
\begin{equation}
m=\frac{\alpha}{2\pi^2}\int
d^2\qpara\frac{m}{q_3^2+(q_4-m)^2+m^2}\int_0^\infty d\qperp^2
\frac{\exp(-q_\perp^2/2|eH|)}{\qpara^2+\qperp^2 +
\mathrm{Re}\Pi(\qpara^2,\qperp^2)}, \label{gap}
\end{equation}
where $\qpara^2 = q_3^2 + q_4^2$.

Before proceeding further we note that, like what has usually been
done in the literature, Eq.~\eqref{gap} can be obtained directly
from Eq.~\eqref{m} by making a Wick rotation to Euclidean space.
While, as remarked in Ref.~\cite{Itzykson}, this procedure is easy
to justify in perturbation theory by neglecting the possibility of
dynamically generated singularities in the first and third
quadrants of the complex energy plane, it is highly nontrivial in
nonperturbative studies and must be performed with care. Our
detailed analysis of the photon polarization in the BVA shows
clearly that no such singularities will be generated dynamically
in the first and third quadrants of the complex $q_0$-plane, thus
leading to a justification for the Wick rotation to Euclidean
space in this problem.

\begin{figure}[t]
\begin{center}
\includegraphics[width=4.0in,keepaspectratio=true,clip=true]{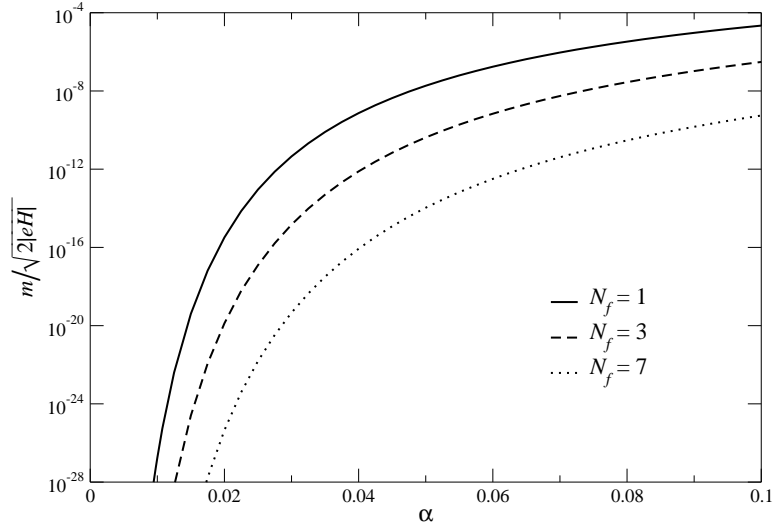}
\end{center}
\caption{Plot of $m$ as a function of $\alpha$ for several values of
$N_f$.} \label{fig:m}
\end{figure}

The generalization of our result to the case of QED with $N_f$
fermion flavors can be done straightforwardly by the replacement
$\Pi(\qpara^2,\qperp^2)\to N_f\,\Pi(\qpara^2,\qperp^2)$ in
Eqs.~\eqref{D}, \eqref{m1} and \eqref{gap}. We have numerically
solved Eq.~\eqref{gap} to obtain $m$ as a function of $\alpha$ for
several values of $N_f$. The results are displayed in
Fig.~\ref{fig:m}. Numerical analysis shows that the solution of
Eq.~\eqref{gap} can be fit by the following analytic expression:
\begin{equation}
m=a\,\sqrt{2|eH|}\;\beta(\alpha)\,
\exp\left[-\frac{\pi}{\alpha\log(b/N_f\,\alpha)}\right],
\label{manalytic}
\end{equation}
where $a$ is a constant of order one, $b\simeq 2.3$, and
$\beta(\alpha)\simeq N_f\,\alpha$.

It is instructive to consider the case of QED with a large number of
fermion flavors, i.e., the $1/N_f$ expansion in large-$N_f$
QED~\cite{Hong:1996pv,Gusynin:2003dz}. The advantage of the $1/N_f$
expansion is that it allows for a calculation of the dynamical
fermion mass without a restriction to small $\alpha$.  Specifically,
we take the limit $N_f\to\infty$ with $N_f\,\alpha$ finite and
fixed. This can be achieved by a rescaling of the coupling constant
$e\to e/\sqrt{N_f}$, which together with Eq.~\eqref{manalytic} leads
to the following result for the dynamical fermion mass in the
large-$N_f$ limit:
\begin{equation}
m\simeq a\;\overline{\alpha}\,\sqrt{2|eH|}\,\exp
\left[-\frac{N_f\,\pi}{\overline{\alpha}\log(b/\overline{\alpha})}\right],
\label{mlargeN}
\end{equation}
where $\overline{\alpha}$ is the rescaled fine-structure
constant, which remains finite and fixed in the limit
$N_f\to\infty$ as noted above. We will argue in the next
section that Eq.~\eqref{mlargeN} can also be obtained from
the 2PI (two-particle-irreducible) effective action
truncated at next-to-leading order in the $1/N_f$
expansion, which resums exactly the same infinite subset of
diagrams as the SD equations truncated in the BVA.

We end this section by showing that the on-shell gauge independence
of the fermion self-energy in the BVA can also be understood in
terms of the wave function renormalization, i.e., the residue at the
fermion pole. Since the gauge dependent contribution as well as
contributions from the terms proportional to $\qpara^\mu
\qpara^\nu/\qpara^2$ in the full photon propagator are proportional
to $(\gammapara \cdot \ppara + m)$, they contribute only to the wave
function renormalization in the full fermion propagator if the
fermion self-energy were evaluated off the fermion mass shell. Hence
they do not affect the position of the fermion pole, which is at
$\ppara^2=-m^2$ and gauge independent. This is precisely the reason
that the physical dynamical fermion mass has to be determined by the
pole of the full fermion propagator in a consistent truncation. The
wave function renormalization is gauge dependent, but this is
perfectly fine since it is not a physical observable. Furthermore,
the wave function renormalization can be absorbed into a
redefinition of the fermion field. This in turn implies that up to a
field redefinition the physical fermion pole at $\ppara^2=-m^2$ has
unit residue.

\section{Comparison to previous works and discussions}
\label{Sec:Comparison}

Our results for the dynamical fermion mass differ from those
obtained in the BVA in
Refs.~\cite{Gusynin:1995gt,Gusynin:1995nb,Leung:1996qy,Lee:1997zj,Gusynin:1998zq,Gusynin:1999pq,Alexandre:2000nz,Alexandre:2001vu,Gusynin:2003dz,Kuznetsov:2002zq},
in which the truncated fermion self-energy were evaluated
exclusively off the fermion mass shell at $\ppara^2=0$.
Consequently, the results for the dynamical fermion mass obtained in
these earlier studies are inevitably gauge dependent and cannot be
identified unambiguously as the physical dynamical fermion mass. We
now argue that to a large extent those earlier results can be
attributed to inconsistent truncation schemes, gauge dependent
artifacts, or both.

In
Refs.~\cite{Gusynin:1995gt,Gusynin:1995nb,Leung:1996qy,Lee:1997zj}
chiral symmetry breaking in a strong magnetic field was first
studied in the so-called rainbow approximation, in which the bare
vertex and the free photon propagator were used. Apart from the
gauge dependence, these earlier results are found to have a very
different functional dependence on the gauge coupling. This is a
consequence of an inconsistent truncation of the SD equations in
that, as displayed in Fig.~\ref{fig:SDBVA}, the photon propagator
that enters the SD equations in the BVA is the full one, hence could
be replaced by the bare one if and only if the vacuum polarization
effects are shown to be negligible. However, this is not the case
for the problem under consideration. As discussed in
Sec.~\ref{Sec:Proof}, there is a strong screening effect in the
longitudinal components of the photon propagator for photons with
momenta $m^2\ll|\qpara^2|\ll|eH|$ and $\qperp^2\ll|eH|$. This is
precisely the reason that the numerical values of the dynamical
fermion mass found in
Refs.~\cite{Gusynin:1995gt,Gusynin:1995nb,Leung:1996qy,Lee:1997zj}
tend to be overestimated by several orders of magnitude when
compared to what we have obtained in Sec.~\ref{Sec:Proof}.

More recently, chiral symmetry breaking in a strong magnetic field
has been studied in the so-called improved rainbow
approximation~\cite{Gusynin:1998zq,Gusynin:1999pq,Gusynin:2003dz,Kuznetsov:2002zq},
in which the bare vertex and the full photon propagator were used.
We note that the improved rainbow approximation used in
Refs.~\cite{Gusynin:1998zq,Gusynin:1999pq,Gusynin:2003dz,Kuznetsov:2002zq}
is exactly the same as the bare vertex approximation used in this
article. It can be verified fairly easily that the truncated SD
equations in both approximations resum identically the same infinite
subset of diagrams.

The authors of
Refs.~\cite{Gusynin:1998zq,Gusynin:1999pq,Gusynin:2003dz} claimed
that (i) in covariant gauges there are one-loop vertex corrections
arising from the term $\qpara^\mu\qpara^\nu/q^2\qpara^2$ in the full
photon propagator that are not suppressed by powers of $\alpha$ (up
to logarithms) and hence need to be accounted for; (ii) there exists
a noncovariant and nonlocal gauge in which, and only in which, the
BVA is a reliable truncation of the SD equations that consistently
resums these one-loop vertex corrections. The gauge independent
analysis in the BVA, as presented in this article, shows clearly
that such contributions vanish identically on the fermion mass
shell. Furthermore, it is evident that if one expands
diagrammatically the SD equations in the BVA, one finds by induction
that to all orders in the loop expansion there is \emph{not} a
single diagram with vertex corrections being resummed by the SD
equations (see Fig.~\ref{fig:SDBVA}). This statement is true in
\emph{any} gauge (be it covariant or otherwise), because gauge
fixing affects only the explicit form of the (free or full) photon
propagator but not the topology of the Feynman diagrams. Therefore,
both the above quoted conclusions are incorrect.

We emphasize that the WT identity is a \emph{necessary} condition for
establishing the gauge independence of the dynamical fermion mass,
but it is far from sufficient. While the WT identity guarantees the truncated
vacuum polarization is transverse, it guarantees \emph{only} the truncated
on-shell fermion self-energy is gauge independent. Hence, the
dynamical fermion mass is gauge independent only when determined by
the position of the fermion pole obtained in a consistent
truncation. This is tantamount to evaluating the truncated fermion
self-energy on the fermion mass shell. Even though the WT identity
in the BVA is verified in Refs.~\cite{Gusynin:1998zq,Gusynin:1999pq}
in a particular noncovariant and nonlocal gauge, this does not
guarantee that the dynamical fermion mass obtained therein from the
truncated fermion self-energy evaluated off the fermion mass shell
will be gauge independent. In fact, the authors of
Refs.~\cite{Gusynin:1998zq,Gusynin:1999pq,Gusynin:2003dz} simply chose a
particular noncovariant and nonlocal gauge such that the gauge dependent
contribution happens to cancel contributions from terms proportional
to $\qpara^\mu \qpara^\nu/\qpara^2$ in the full photon
propagator~\cite{MiranskyPC}. Our gauge independent analysis in the
BVA reveals clearly that such a gauge fixing not only is ad hoc and
unnecessary, but also leaves the issue of gauge independence
unaddressed.

In Ref.~\cite{Kuznetsov:2002zq} its authors claimed that (i) in QED
with $N_f$ fermion flavors a critical number $N_{cr}$ exists for any
value of $\alpha$, such that chiral symmetry remains unbroken for
$N_f >N_{cr}$; (ii) the dynamical fermion mass is generated with a
double splitting for $N_f< N_{cr}$. As can be gleaned clearly from
Fig.~\ref{fig:m} or from the large-$N_f$ result,
Eq.~\eqref{mlargeN}, both the conclusions quoted above are
incorrect. They are gauge dependent artifacts of an inconsistent
truncation: On the one hand, the SD equation for the momentum
independent fermion self-energy was obtained (in an unspecified
``appropriate'' gauge) in the BVA \emph{within} the LLLA. On the
other hand, the vacuum polarization was calculated in the BVA but
\emph{beyond} the LLLA. This however is \emph{not} a consistent
truncation of the SD equations since, as shown in the end of
Sec.~\ref{Sec:SD}, the WT identity in the BVA can be satisfied only
within the LLLA.

The results of Ref.~\cite{Kuznetsov:2002zq} suggest that in the
inconsistent truncation as well as in the unspecified
``appropriate'' gauge used there, the unphysical, gauge dependent
contributions from higher Landau levels are so large that they
become dominant over the physical, gauge independent contribution
from the LLL and lead to the authors' incorrect conclusions. Hence,
we emphasize that the LLL dominance in a strong magnetic field
should be understood in the context of consistent truncation schemes
as follows. Contributions to the dynamical fermion mass from higher
Landau levels that are obtained in a (yet to be determined)
consistent truncation of the SD equations are subleading when
compared to that from the LLL obtained in the consistent BVA
truncation.  Our analysis of the WT identity in the BVA but beyond
the LLLA indicates that in order to go beyond the LLLA, one has to
use truncation schemes that consistently account for vertex
corrections.  To the best of our knowledge, consistent truncations
of the SD equations that include vertex corrections either within or
beyond the LLLA have not appeared in the literature. In view of
this, a detailed study of the general structure as well as the
analytic properties of the full vertex akin to that in
Ref.~\cite{Ball:1980ay} but in the presence of a constant external
magnetic field is of crucial importance.

Extension to a momentum dependent fermion self-energy in the BVA has
been studied in a certain gauge by neglecting corrections to the
vacuum
polarization~\cite{Gusynin:1998zq,Gusynin:1999pq,Alexandre:2000nz,Alexandre:2001vu,Gusynin:2003dz}.
It was found that the fermion self-energy has very weak momentum
dependence in the region relevant to fermion pairing in a strong
magnetic field. Since a momentum dependent fermion self-energy
cannot fulfill the WT identity in the BVA, the results obtained in
these studies are inevitably gauge dependent. It is, however,
important to note that when compared to the corresponding results
obtained using a momentum independent fermion self-energy, these
results lend support to the reliability of a momentum independent
fermion self-energy and, consequently, of the WT identity in the BVA
in the momentum region $|\ppara^2|\ll|eH|$ that is relevant to the
dynamics of fermion pairing in a strong magnetic field. Similar
study has also been carried out in a certain gauge in the case of
QED with an extra charge neutral, self-interacting scalar field that
couples to the fermion through the Yukawa
interaction~\cite{Ferrer:2000ed,Elizalde:2002ca}.

Despite obtained from gauge dependent calculations in a certain
gauge, as will be argued momentarily, the conclusion from these
studies~\cite{Gusynin:1998zq,Gusynin:1999pq,Alexandre:2000nz,Alexandre:2001vu,Gusynin:2003dz}
that a momentum independent fermion self-energy or, equivalently,
the WT identity in the BVA is reliable in the momentum region
relevant to chiral symmetry breaking in a strong magnetic field
remains valid. In fact, this is precisely the underlying physical
reason that the BVA is a consistent truncation of the SD equations
in the LLLA.

In the BVA the presence of momentum dependent terms in the fermion
self-energy leads to violation of the WT identity, hence introduces
gauge dependent contributions to the SD equations even though the
latter are evaluated on the fermion mass shell. We now argue that
such gauge dependent contributions are of higher orders when
compared to the dynamical fermion mass obtained from a momentum
independent fermion self-energy. Therefore, within the LLLA the
dynamical fermion mass can be reliably calculated in the BVA by
using a momentum independent fermion self-energy.

To begin with, we first note that the infinite subset of diagrams
being resummed in
Refs.~\cite{Gusynin:1998zq,Gusynin:1999pq,Alexandre:2000nz,Alexandre:2001vu,Gusynin:2003dz}
are identical to, as well as in the same kinematic region as, those
in the present article. The only difference is that in those
references the calculation is carried out in a certain gauge and the
fermion self-energy is evaluated off the fermion mass shell at
$\ppara^2=0$. Since the dynamical fermion mass is much less than the
Landau energy, i.e., $m^2\ll|eH|$, the LLL fermion self-energy
$\Sigma(\ppara)$ with momentum $|\ppara^2|\ll|eH|$ can be expanded
in powers of $\ppara^2/|eH|$ around the fermion mass shell in the
BVA (i.e., $\ppara^2=-m^2$) as [see Eq.~\eqref{SigmaComp}]
\begin{align}
A_\parallel(\ppara)&=a(\xi)\,x+\mathcal{O}(x^2),\\
B(\ppara)&=m\,[1+b(\xi)\,x+\mathcal{O}(x^2)],\label{ApBp}
\end{align}
where $x=(\ppara^2+m^2)/|eH|$, and $a(\xi)$ and $b(\xi)$ are
$\xi$-dependent coefficients presumably of order one. This, together
with the \emph{gauge dependent} results obtained in the
LLLA~\cite{Gusynin:1998zq,Gusynin:1999pq,Alexandre:2000nz,Alexandre:2001vu,Gusynin:2003dz}
regarding the reliability of a momentum independent fermion
self-energy in the BVA in the momentum region $|\ppara^2|\ll|eH|$,
indicates clearly that the gauge dependent contributions to the
dynamical fermion mass arising from the momentum dependent terms in
$A_\parallel(\ppara)$ and $B(\ppara)$ are of order
$|\ppara^2|/|eH|$. Hence physical quantities sensitive only to the
kinematic region of momenta $|\ppara^2|\ll|eH|$ can be reliably
calculated in the BVA, and are gauge independent up to corrections
of order $|\ppara^2|/|eH|\ll 1$. This in turn means that, consistent
with the underlying assumption of the LLLA, the kinematic regime of
low-energy photon exchange ($|\qpara^2|,\qperp^2\ll|eH|$) is
precisely that which determines the dynamics of fermion pairing in a
strong magnetic field.

We are now in a position to establish direct contact between the SD
equations truncated in the BVA and the 2PI
(two-particle-irreducible) effective action truncated at the lowest
nontrivial order in the loop expansion as well as in the $1/N_f$
expansion. Since the 2PI effective action is constructed in terms of
the full propagators and bare vertices, the corresponding WT
identities usually cannot be satisfied. In general, any truncation
of the 2PI effective action is inevitably gauge dependent and hence,
strictly speaking, an inconsistent one. Nevertheless, it has been
shown recently in
Refs.~\cite{Arrizabalaga:2002hn,Mottola:2003vx,Carrington:2003ut}
that the truncated \emph{on-shell} 2PI effective action has a
controlled gauge dependence with the explicit gauge dependent terms
always appearing at higher order. This is reminiscent of the
situation that the dynamical fermion mass obtained from the on-shell
SD equations truncated in the BVA with a momentum dependent fermion
self-energy that violates the corresponding WT identity, is gauge
independent up to corrections of higher order

In Appendix~\ref{Appendix:2PI} we briefly summarize some exact
relations derived from the 2PI effective action that are useful to
our discussion here. The interested reader is referred to the
literature~\cite{Cornwall:1974vz} for further details. For QED in a
constant external magnetic field, the 2PI effective action is given
by
\begin{align}
\Gamma[A,G,\mathcal{D}]&=S_0[A]-i\mathrm{Tr}\log
G^{-1}-i\mathrm{Tr}\,S^{-1}(G-S)\nn\\
&\quad+\frac{i}{2}\mathrm{Tr}\log\mathcal{D}^{-1}
+\frac{i}{2}\mathrm{Tr}\,D^{-1}(\mathcal{D}-D)+\Gamma_2[G,\mathcal{D}],
\end{align}
where $S_0[A]$ is the classical Maxwell action with $A_\mu$ being
the external gauge field, and the propagators are the same as those
in Sec.~\ref{Sec:SD}. Specifically, we consider the LLLA hence the
fermion propagators are those for the LLL fermions. In actual
applications, the 2PI effective action is truncated at some order in
a chosen expansion parameter. For the purpose of our discussion, it
suffices to consider only the lowest nontrivial order truncation,
namely, at two-loop order in the loop expansion in QED or at
next-to-leading order [i.e., $\mathcal{O}(N_f^0)$] in the $1/N_f$
expansion in large-$N_f$ QED.

\begin{figure}[t]
\begin{center}
\includegraphics[width=3.5in,keepaspectratio=true,clip=true]{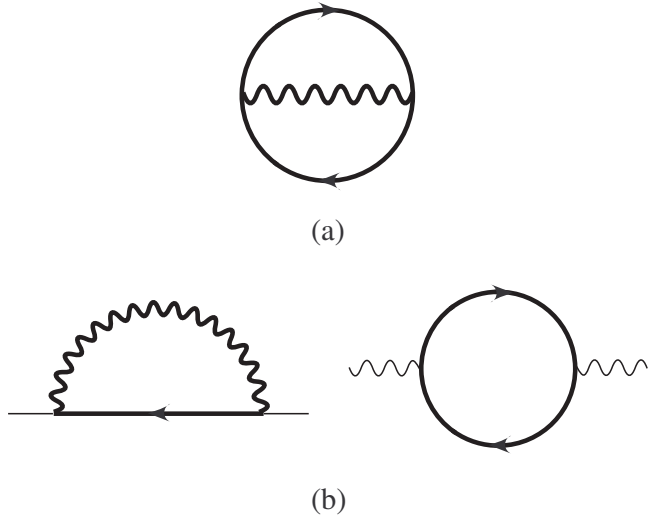}
\end{center}
\caption{(a) Contribution to the 2PI effective action at two-loop
order in the loop expansion or at next-to-leading order in the
$1/N_f$ expansion. (b) The corresponding fermion self-energy and
vacuum polarization at the same order. All internal lines denote the
full propagators and all vertices are the bare ones.}
\label{fig:2PI}
\end{figure}

At the lowest nontrivial order in the loop expansion or in the
$1/N_f$ expansion $\Gamma_2[G,\mathcal{D}]$ is given by
\begin{align}
\Gamma_2[G,\mathcal{D}]&=-\frac{ie^2}{2}\int d^4x\,d^4y\,
\mathrm{tr}\,\gamma^\mu G(x,y)\gamma^\nu
G(y,x)\mathcal{D}_{\mu\nu}(x,y),
\end{align}
where in the $1/N_f$ expansion $e$ is understood as the rescaled
coupling that remains finite and fixed in the limit $N_f\to\infty$.
Functional differentiations of $\Gamma[A,G,\mathcal{D}]$ with
respect to the full propagators in the absence of sources leads to
the SD equations for the respective propagators. The corresponding
fermion self-energy and vacuum polarization are obtained by varying
$\Gamma_2[G,\mathcal{D}]$ with respect to the corresponding full
propagators. The resultant SD equations are found to be exactly the
same as those in the BVA with the corresponding fermion self-energy
and vacuum polarization given by Eqs.~\eqref{Sigmaxy} and
\eqref{Pixy}, respectively. A diagrammatical representation of the
2PI effective action is depicted in Fig.~\ref{fig:2PI}, in which it
is noted that the vertices that appear in $\Gamma_2[G,\mathcal{D}]$
as well as in the corresponding fermion self-energy and vacuum
polarization are the \emph{bare} ones. With the above observation
one can conclude fairly easily that the 2PI effective action
truncated at the lowest nontrivial order in the loop expansion or in
the $1/N_f$ expansion resums identically the same infinite subset of
diagrams as the SD equations truncated in the BVA.

The direct correspondence with the 2PI effective action truncated at
the lowest nontrivial order has the following important
consequences. On the one hand, the WT identity in the BVA \eqref{WT}
guarantees that the truncation at the lowest nontrivial order in the
loop expansion or in the $1/N_f$ expansion with a momentum
independent fermion self-energy is a consistent truncation of the
2PI effective action in the momentum region relevant to chiral
symmetry breaking in a strong magnetic field. On the other hand, the
facts that vertex corrections in the 2PI effective action start to
appear at the next-to-lowest-nontrivial order in the loop expansion
or in the $1/N_f$ expansion and that the truncated on-shell 2PI
effective action (in general with a momentum dependent fermion
self-energy) has a controlled gauge dependence with the explicit
gauge dependent terms always appearing at higher order, provide an
unequivocal justification of our argument that in a strong magnetic
field the dynamical fermion mass to leading order in the gauge
coupling or in the $1/N_f$ expansion can be reliably calculated in
the BVA, and is gauge independent up to corrections of higher order.
Whereas a direct verification by explicit calculation is
indispensable, to the best of our knowledge consistent truncations
of the SD equations that include vertex corrections either within or
beyond the LLLA have not appeared in the literature. Clearly such a
task is beyond the scope of the current article, and will be the
subject of further investigations.

\section{Summary and conclusions}\label{Sec:Conclusions}

In this article we presented a critical and detailed study of chiral
symmetry breaking in QED in a constant external magnetic field. The
main goal is to determine in a gauge independent manner the
dynamical fermion mass generated through chiral symmetry breaking.

We focused on the strong field limit as well as on the weak coupling
regime. The former leads to a wide separation of energy scales such
that the dynamics of fermion pairing is dominated by the lowest
Landau level. The latter allows for a controlled, nonperturbative
calculation of the dynamical fermion mass, in the sense that there
exists a systematic expansion in powers of the gauge coupling (up to
logarithms) such that contributions of leading order in the gauge
coupling that arise form every order in the loop expansion can be
consistently accounted for, while subleading contributions are
suppressed by powers of the gauge coupling (up to logarithms).
Consequently, in the weak coupling regime a gauge independent,
consistent truncation of the SD equations is tantamount to a good
and reliable approximation.

The WT identity in the BVA is at the heart of our proof that the BVA
is a consistent truncation of the SD equations within the LLLA.  We
first verified that, with the corrected orthonormal condition of the
$E_p$ functions for the LLL fermions, the conclusion of
Ref.~\cite{Ferrer:1998vw} that in order to satisfy the WT identity
in the BVA within the LLLA, the fermion self-energy has to be a
momentum independent constant remains valid. Furthermore, we showed
that the WT identity in the BVA can be satisfied only within the
LLLA.  The proof then proceeds with the assumption that the BVA is a
consistent truncation such that the position of the fermion pole
obtained therein is gauge independent. With this assumption, the WT
identity in the BVA implies that the pole of the LLL fermion
propagator is located at $\ppara^2=-m^2$, with $m$ being the
momentum independent fermion self-energy \emph{as well as} the
physical, gauge independent dynamical fermion mass. The proof is
completed by verifying that such an assumption does not lead to any
inconsistency. This is achieved by showing that (i) the truncated
vacuum polarization is transverse, (ii) the truncated fermion
self-energy evaluated on the fermion mass shell is manifestly gauge
independent. In particular, we showed that the would-be gauge
dependent contribution to the truncated fermion self-energy that
arises from the gauge dependent term in the full photon propagator
vanishes identically on the fermion mass shell. This detailed study
leads to a gauge independent, nonperturbative calculation of the
dynamical fermion mass to leading order in the gauge coupling.
Furthermore, it allows one to identify unambiguously the infinite
subset of diagrams that contribute to chiral symmetry breaking in a
strong magnetic field.

We made a detailed comparison between our results and those obtained
in previous
works~\cite{Gusynin:1995gt,Gusynin:1995nb,Leung:1996qy,Lee:1997zj,Gusynin:1998zq,Gusynin:1999pq,Alexandre:2000nz,Alexandre:2001vu,Gusynin:2003dz,Kuznetsov:2002zq}.
The gauge independent analysis we presented in this article shows
clearly that the results as well as the conclusions of those earlier
studies are inevitably gauge dependent and can be attributed to
inconsistent truncation schemes, gauge dependent artifacts, or both.
Nevertheless, based on the gauge dependent
results~\cite{Gusynin:1998zq,Gusynin:1999pq,Alexandre:2000nz,Alexandre:2001vu,Gusynin:2003dz}
obtained within the LLLA regarding the reliability of a momentum
independent fermion self-energy in the BVA in the momentum region
$|\ppara^2|\ll|eH|$, we argued that in a strong magnetic field the
dynamical fermion mass can be reliably calculated in the BVA and is
gauge independent up to corrections of higher order. This is
consistent with the underlying assumption of the LLLA that the
kinematic region of low-energy photon exchange
($|\qpara^2|,\qperp^2\ll|eH|$) is precisely that which determines
the dynamics of fermion pairing in a strong magnetic field.

Motivated by the fact that the gauge dependence of the dynamical
fermion mass calculated in the BVA but with a momentum dependent
fermion self-energy is subleading, we established a direct contact
between the SD equations truncated in the BVA and the 2PI effective
action truncated at the lowest nontrivial order in the loop
expansion as well as in the $1/N_f$ expansion. This direct
correspondence, together with the facts that vertex corrections in
the 2PI effective action start to appear at the
next-to-lowest-nontrivial order in the loop expansion or in the
$1/N_f$ expansion and that the truncated on-shell 2PI effective
action has a controlled gauge dependence with the explicit gauge
dependent terms always appearing at higher order, provides the
justification of our argument that in a strong magnetic field the
dynamical fermion mass can be reliably calculated in the BVA to
leading order in the gauge coupling or in the $1/N_f$ expansion.

In conclusion, the presence of a strong, constant external magnetic
field provides a consistent truncation of the Schwinger-Dyson
equations in terms of the bare vertex approximation within the
lowest Landau level approximation. This allows us to study in a
gauge independent manner the physics of chiral symmetry breaking in
QED in an external magnetic field and to obtain a solution for the
physical dynamical fermion mass. The gauge independent approach to
the Schwinger-Dyson equations in gauge theories that we presented in
this article is quite general in nature, and hence not specific to
the problem of chiral symmetry breaking in a strong magnetic field
discussed here. We believe this approach will be useful in other
areas of physics that also require a nonperturbative understanding
of gauge theories, such as the studies of hadronic structure and the
connection between quark confinement and chiral symmetry breaking in
nuclear and particle physics~\cite{Roberts:1994dr,Alkofer:2000wg},
the physics of superconductivity and superfluidity in condensed
matter systems~\cite{Pereg,Calzetta:2005dt}, as well as the
equilibrium and nonequilibrium properties of hot and dense
matter~\cite{Rebhan:2001wt,Berges:2004pu}.

\section*{Acknowledgments}

We thank P.\ Jasinski for bringing our attention to an algebraic
oversight in Refs.~\cite{Leung:1996qy,Lee:1997zj}. We also thank D.\
Boyanovsky and E.\ Mottola for useful discussions. S.-Y.W.\ would
like to thank the hospitality of the Theoretical Division of the Los
Alamos National Laboratory, where part of this article was written.
C.N.L.\ thanks the Institute of Physics at the Academia Sinica in
Taiwan for its hospitality and support during the writing of the
manuscript. This work was supported in part by the U.S.\ Department
of Energy under grant DE-FG02-84ER40163 and by the National Science
Council of Taiwan under grant NSC-95-2112-M-032-010.

\appendix

\section{Vacuum current in a constant magnetic field}\label{Appendix:VC}

Using the Ritus $E_p$ functions and the general
structure of the fermion self-energy, we shall show in a
\emph{nonperturbative} manner that the vacuum current vanishes
identically in a constant external magnetic field. The vacuum
current is defined to be
\begin{align}
J^\mu&=ie\,\mathrm{tr}\,\gamma^\mu G(x,x)\nn\\
&=ie\,\mathrm{tr}\sumint\,\frac{d^4p}{(2\pi)^4}\,\overline{E}_p(x)\,\gamma^\mu
E_p(x)\,\frac{1}{\gamma\cdot p+\Sigma(p)},
\end{align}
where in the second equality we have used the momentum
representation of the full fermion propagator given by
Eq.~\eqref{GX}. It is noted that in this appendix the fermion
propagator and the fermion self-energy will \emph{not} be
restricted to those for the LLL fermions.

With the properties \eqref{DeltaProp} and \eqref{Drho}, the
integration over $p_2$, one of the spin sums, as well as the trace
can be done easily, leading to
\begin{align}
J^\mu&=\frac{ie}{\pi}\,|eH|\sum^\infty_{l=0}\sideset{}{'}\sum_{\sigma=\pm
1}\int\frac{d^2\ppara}{(2\pi)^2}\frac{[1+A_\parallel(p)]\,\ppara^\mu}
{[1+A_\parallel(p)]^2\,\ppara^2+[1+A_\perp(p)]^2\,\pperp^2+B^2(p)},
\end{align}
where $\ppara^2=-p_0^2+p_3^2$, $\pperp^2=2|eH|l$, and use has been
made of the Dirac matrix structure of the fermion self-energy in
Eq.~\eqref{SigmaComp}. Since $A_{\parallel,\perp}(p)$ and $B(p)$ are
functions of the longitudinal and transverse momentum squared, one
finds that $J^\mu$ vanishes identically. The same conclusion also
applies to the vacuum current in the LLLA, in which the sum over $l$
is restricted to the $l=0$ term and that over $\sigma$ is restricted
to $\sigma=\sgn(eH)$.

We note that the vanishing of the vacuum current in a constant
external magnetic field is a consequence of the Lorentz boost
invariance along, as well as the rotational invariance about, the
direction of the constant external magnetic field, namely, the
$x_3$-direction.

\section{2PI effective action}\label{Appendix:2PI}

In this appendix we briefly summarize some exact relations derived
from the 2PI (two-particle-irreducible) effective
action~\cite{Cornwall:1974vz} that are useful to our discussion. For
presentational simplicity we consider here only bosonic fields, the
generalization to fermionic fields is straightforward. The
generating functional for connected correlation functions is defined
as
\begin{align}
Z[J,K]&=e^{iW[J,K]}\nn\\
&=\int\mathcal{D}\varphi\,
e^{i(S[\varphi]+J_i\varphi^i+\frac{1}{2}\varphi^i
K_{ij}\varphi^j)},\label{Z}
\end{align}
where $S[\varphi]$ is the classical action, $\varphi^i$ denotes
bosonic fields and $J$ ($K$) is the auxiliary local (bilocal)
source. We use a shorthand notation in which Latin indices stand
for space-time variables as well as internal indices, and
summation and integration over repeated indices are understood,
e.g.,
\begin{equation}
J_i\varphi^i=\int d^4x\,J(x)\,\varphi(x).
\end{equation}

The mean field $\phi^i=\langle\varphi^i\rangle$ and the
\emph{connected} two-point function $G^{ij}=\langle
T\varphi^i\varphi^j\rangle-\langle\varphi^i\rangle
\langle\varphi^j\rangle$ are obtained by functional differentiations
of $W[J,K]$ with respect to the local source $J$ as
\begin{equation}
i\frac{\delta W}{\delta(iJ_i)}=\phi^i,\qquad i\frac{\delta^2
W}{\delta(iJ_i)\delta(iJ_j)}=G^{ij}.\label{meanfield}
\end{equation}
Functional differentiations of $W[J,K]$ with respect to the
bilocal source $K$ may generate also disconnected correlation
functions. For example, differentiating once with respect to $K$
leads to
\begin{equation}
i\frac{\delta W}{\delta(iK_{ij})}=
\frac{1}{2}(\phi^i\phi^j+G^{ij}). \label{functionalw}
\end{equation}
A functional Legendre transform in the mean field $\phi$ and the
connected two-point function $G$ leads to the 2PI effective action
\begin{equation}
\Gamma[\phi,G]=W[J,K]-J_i\phi^i-\frac{1}{2}K_{ij}
(\phi^i\phi^j+G^{ij}).\label{legendre}
\end{equation}
From Eqs.~\eqref{meanfield} and \eqref{functionalw} one can derive
the relations
\begin{equation}
\frac{\delta \Gamma[\phi,G]}{\delta \phi^i}=-J_i-K_{ij}\phi^j,
\quad\frac{\delta \Gamma[\phi,G]}{\delta G^{ij}}=-\frac{1}{2}
K_{ij}.\label{JK}
\end{equation}
The 2PI effective action can be cast into a very convenient form
which has a simple diagrammatical interpretation in terms of 2PI
diagrams (for fermionic fields the factors of $1/2$ before the trace
terms are replaced by $-1$)
\begin{align}
\Gamma[\phi,G]&=S_{0}[\phi]+\frac{i}{2}\mathrm{Tr}\log G^{-1}
+\frac{i}{2}\mathrm{Tr}\,G_0^{-1}(G-G_0)+\Gamma_2[\phi,G],\label{2PI}
\end{align}
where $S_0$ is the free part of the classical action and $G_0$ is
the bare two-point function
\begin{equation}
G_{0ij}=\biggl(-i\frac{\delta^2 S_0[\varphi]}{\delta\varphi^i
\delta\varphi^j}\biggr)^{-1}.
\end{equation}
The functional $\Gamma_2[\phi,G]$ is the sum of all 2PI skeleton
vacuum diagrams with \emph{bare} vertices and \emph{full}
propagators. Here, skeleton diagrams are those without self-energy
insertions.

The equations of motion for the mean field $\phi$ and the connected
two-point function $G$ are determined by the stationary condition,
i.e., from the implicit functional equation \eqref{JK} for vanishing
sources $J$ and $K$, as
\begin{equation}
\frac{\delta\Gamma[\phi,G]}{\delta\phi^i}=0,\qquad
\frac{\delta\Gamma[\phi,G]}{\delta G^{ij}}=0. \label{EOM}
\end{equation}
With Eq.~\eqref{2PI} the equation of motion for the connected
two-point function in Eq.~\eqref{EOM} becomes
\begin{equation}
G_{ij}^{-1}=G_{0ij}^{-1}+\Sigma_{ij},\label{EOMG}
\end{equation}
where $\Sigma$ is the one-particle-irreducible self-energy
(for fermionic fields the factor of $-2$ is replaced by $1$)
\begin{equation}
\Sigma_{ij}=-2i\frac{\delta\Gamma_2[\phi,G]}{\delta
G^{ij}}.\label{Sigmaij}
\end{equation}
It is noted that Eq.~\eqref{EOMG} is the Schwinger-Dyson equation
for the full propagator $G$.

\end{document}